\documentclass[10pt,twocolumn,letterpaper]{article}

\usepackage[pagenumbers]{cvpr} 

\usepackage{graphicx}
\usepackage{amsmath}
\usepackage{amssymb}
\usepackage{booktabs}
\usepackage{upgreek}
\usepackage{enumitem}

\usepackage[pagebackref,breaklinks,colorlinks]{hyperref}

\usepackage{xcolor}

\usepackage[capitalize]{cleveref}
\crefname{section}{Sec.}{Secs.}
\Crefname{section}{Section}{Sections}
\Crefname{table}{Table}{Tables}
\crefname{table}{Tab.}{Tabs.}

\def\cvprPaperID{4625} 
\def\confName{CVPR}
\def\confYear{2023}

\begin{document}

\title{MiShape: 3D Shape Modelling of Mitochondria in Microscopy}

\author{
    {Abhinanda R. Punnakkal}$^{1}$  \qquad
    {Suyog S.  Jadhav}$^{1}$ \qquad 
    {Alexander Horsch}$^{1}$ \qquad \\
     {\tt\small abhinanda.r.punnakkal@uit.no \qquad
         suyog.s.jadhav@uit.no \qquad
         alexander.horsch@uit.no
         } \\
    {Krishna Agarwal}$^{1}$ \qquad 
    {Dilip K. Prasad}$^{1}$ \qquad \\
    {\tt\small krishna.agarwal@uit.no \qquad 
        dilip.prasad@uit.no} \\
    \small{$^1$UiT The Arctic University of Norway, Troms\o, Norway} \\
}
\date{}
\maketitle

\begin{abstract}
Fluorescence microscopy is a quintessential tool for observing cells and understanding the underlying mechanisms of life-sustaining processes of all living organisms. 
The problem of extracting 3D shape of mitochondria from fluorescence microscopy images remains unsolved 
due to the complex and varied shapes expressed by mitochondria and the poor resolving capacity of these microscopes. 
We propose an approach to bridge this gap by learning a shape prior for mitochondria termed as MiShape, by leveraging high-resolution electron microscopy data.
MiShape is a generative model learned using implicit representations of mitochondrial shapes. It provides a shape distribution that can be used to generate infinite realistic mitochondrial shapes. 
We demonstrate the representation power of MiShape and its utility for 3D shape reconstruction given a single 2D fluorescence image or a small 3D stack of 2D slices. 
We also showcase applications of our method by deriving simulated fluorescence microscope datasets that have realistic 3D ground truths for the problem of 2D segmentation and microscope-to-microscope transformation.
\end{abstract}


\section{Introduction}

The potential of computer vision remains untapped in scientific applications. We are interested in analyzing microscopy data, particularly fluorescence microscopy images of mitochondria. Mitochondria is a cell organelle of utmost importance as it is responsible for the production of most of the chemical energy required for the cell. Understanding the morphology and dynamics of mitochondria serves as looking glass for cell health. Mitochondrial morphological disorders are correlated to several diseases such as Alzheimer's, Parkinson's and Huntington diseases \cite{harmuth2018mitochondrial} and cardiovascular diseases \cite{ong2010mitochondrial} among many others \cite{chinnery2021primary}.

Mitochondria are pleomorphic 3D structures whose size and shape vary greatly with cell types \cite{miyazono2018uncoupled,Conrad2022}. Optical microscopes, specifically fluorescent microscopes are the tools often used by biologists for mitochondrial morphological studies for the following reasons: First, they give high throughput compared to other imaging techniques like Electron Microscopy (EM), and require less time and effort during sample preparation. Second, fluorescent microscopy allows live cell imaging which is not possible using EM. The cell is fairly transparent to white light and it can appear chaotic to human eyes trying to distinguish the different structures in the cell. Fluorescence microscopy allows the labeling of specific structures of interest and provides the contrast needed for the human eye \cite{lichtman2005fluorescence}. 

\begin{figure}[t]
  \centering
\includegraphics[width=\linewidth]{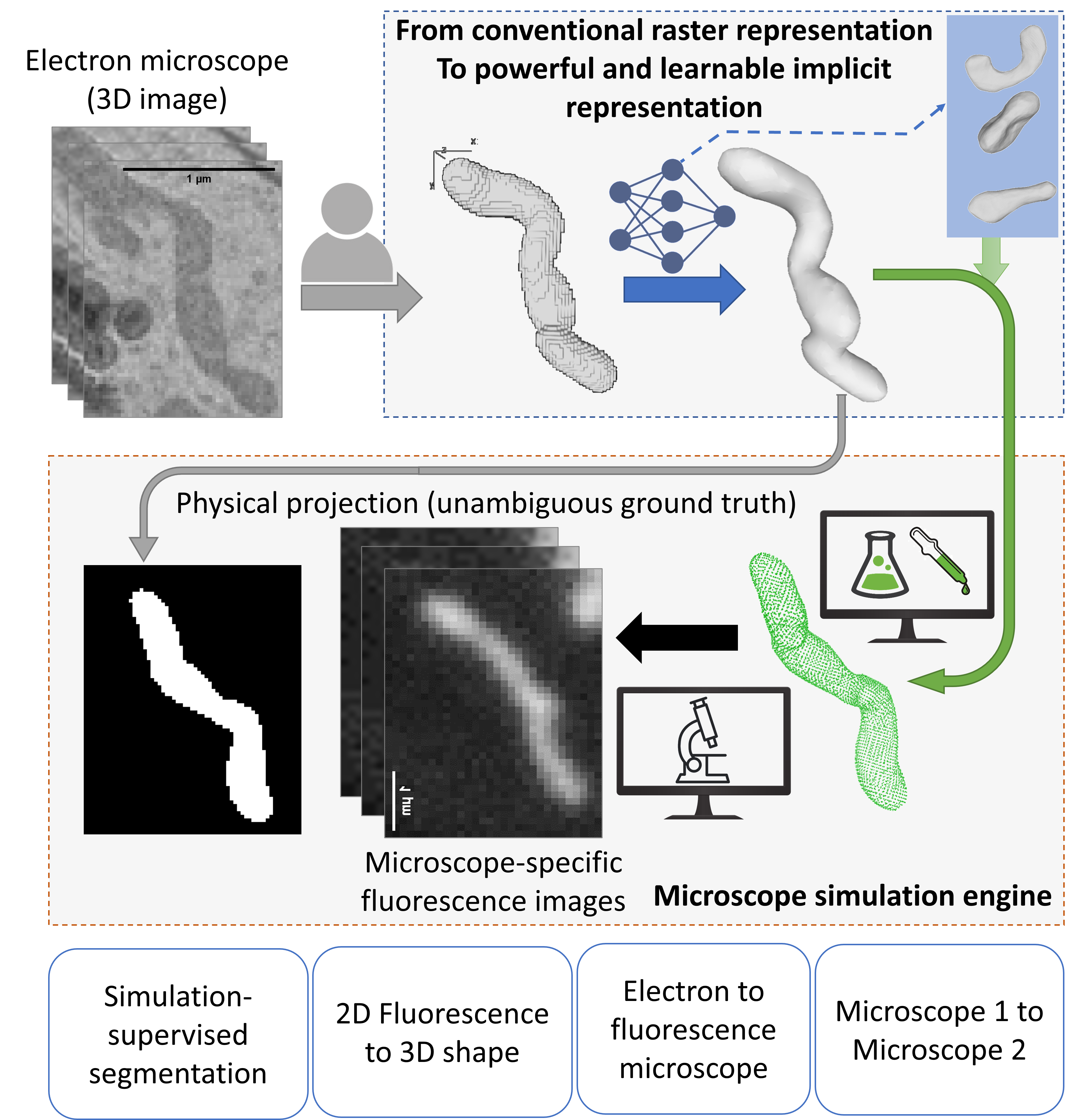}
\caption{We present MiShape, a shape prior for mitochondrial shape. We train MiShape using implicit representations of shapes derived from 3D EM images of mitochondria. MiShape can be used for a variety of paradigms in fluorescence microscopy with interesting applications.}
   \label{fig:teaser}
\end{figure}

\begin{figure*}[t]
  \centering
\includegraphics[width=0.9\linewidth]{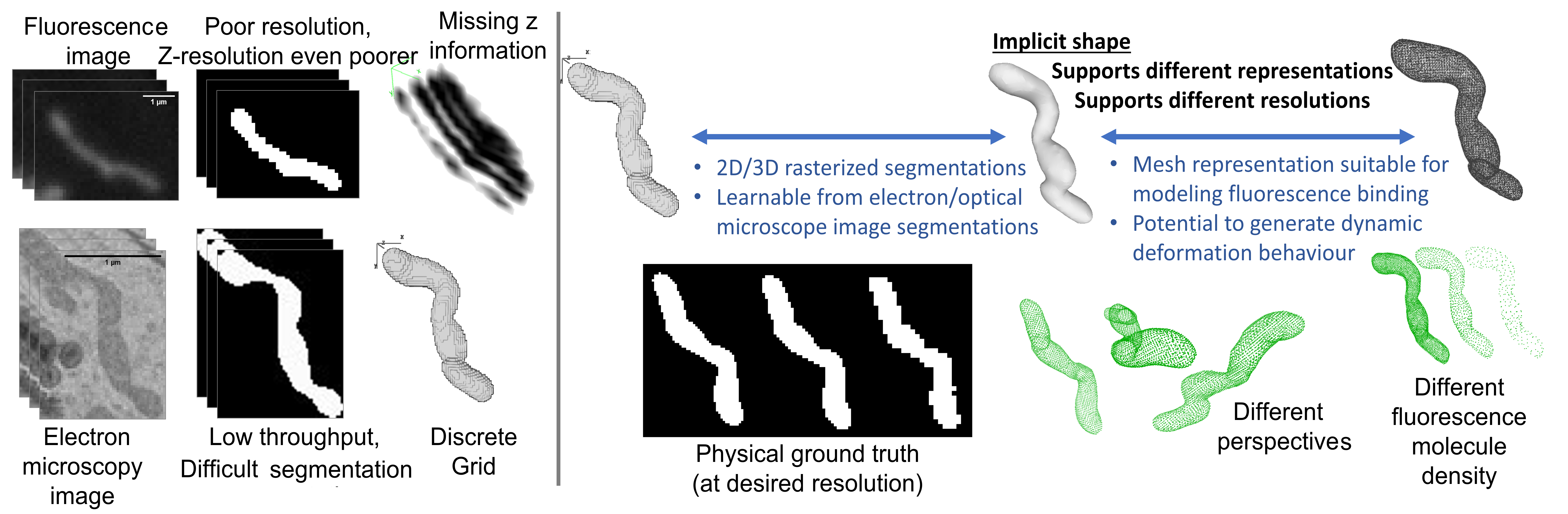}
   \caption{\textbf{Problems with raster representation (left):} Challenges of extracting 3D shape from microscope images are illustrated. Discrete grid in EM images pose challenge in translation of shapes to fluorescence microscope due to different digital and optical resolution. \textbf{Benefits of implicit representation (right):} Implicit shape representation presents several advantages over raster formats, and still supports both raster and mesh formats for flexibly introducing desired properties to fluorescence microscopy images.}
   \label{fig:teaser2}
   \vspace{-5 mm}
\end{figure*}

The resolution of a microscope is the minimum distance at which two distinct points of a specimen can still be seen as separate points. The main disadvantage of using a fluorescence microscope is its limited resolution typically around $200-250\,nm$ which is comparable to the size of mitochondria. Moreover, the resolution in the fluorescence microscope is further poorer in the axial dimension, typically around $500\,nm$. EM is a better alternative in terms of resolution but does not allow live-cell imaging and high throughput. 
Hence, the accurate estimation of 3D shape of mitochondria from fluorescent microscope images remains an unsolved problem in the biological domain.

While 2D segmentation and analytics of mitochondria has garnered interest in both microscopy and deep learning communities, the conventional solutions use parametric representation of shapes of mitochondria \cite{Sekh2021, viana2015quantifying} or morphological operations, it has also been demonstrated in multiple sources that mitochondrial morphologies are far more complex and diverse \cite{miyazono2018uncoupled, Parlakgul2022, xiao2018automatic}. The only solutions currently available involve performing segmentation on 3D EM images \cite{miyazono2018uncoupled,Parlakgul2022,xiao2018automatic,abdollahzadeh2021deepacson,Conrad2022,wei2020mitoem} manually or through semi-supervised learning. Also, the use of these shapes in fluorescence microscopy has not been explored.

We argue that the core obstacle arises from 
rasterized representations, which restricts the mapping between the actual shapes of mitochondria and how they appear in EM and fluorescence images, as illustrated in Fig. \ref{fig:teaser2} (left). Fluorescence microscopy images are a function of several parameters like resolution and magnification that vary from one experiment to another. Using a voxel-based representation \cite{3dgan}  ties us to learning the shape for only one fluorescence experiment setting and does not allow the adaptations for other microscopes and microscope settings. It also enforces digitization at scales comparable to the shapes.  

We resolve the obstacle and propose a computer vision solution for learning and using the shape of mitochondria in a powerful manner for diverse problems involving fluorescence images of mitochondria. Through this work, we enable for the first time learning of sub-cellular organelle's (here mitochondria) shape distribution at nanoscale and use it as an enabler for fluorescence image to 3D shape reconstruction, fluorescence microscope-to-microscope transformation, and EM-to-fluorescence image generation.

We base our work on implicit shape representation, the first such representation in our knowledge for mitochondria or even nanoscale sub-cellular organelles. Previous works \cite{Occupancy_Networks, michalkiewicz1901deep, Chen_2019_CVPR} have illustrated shape learning for everyday world objects. Further, \cite{Yang_2022_CVPR, raju2022deep} developed the method for modelling shapes of human organs, demonstrating that implicit representations derived from rasterized shapes are a good candidate to learn shape distributions. Implicit representation presents several advantages and accords flexibility in simulating fluorescence images, as illustrated in Fig. \ref{fig:teaser2}.

We deliver the advantages of implicit shape representation to the problem of using fluorescence imaging for mitochondrial analysis by defining interesting CV problems and presenting solutions based on implicit representations. This is made possible by using high resolution EM data and implicit representations to generate diverse simulated fluorescence microscopy datasets and their 2D/3D ground truth annotations for different microscopes, resolutions, perspectives of mitochondria, fluorescent molecule densities, and so on. 

Our contributions are summarized as,
\begin{itemize}[leftmargin=*,noitemsep,nolistsep]
    \item Mishape, a generative model for learning the shape of mitochondria using implicit representations. 
    \item Seamless adaptation of shape priors to multiple resolutions and different microscopes for simulating fluorescence images with unambiguous ground truth.
    \item Reconstruction of 3D shapes from a single 2D fluorescence microscopy image. Additionally, 3-slice fluorescence z-stack to 3D shape reconstruction. 
    \item Fluorescence microscope-to-microscope transformation (supplementary).
\end{itemize}
\section{Related Works}




\paragraph{Shape Representation - general and in medical domain:}
The problem of generating realistic 3D shapes has been attempted from different approaches; \cite{3dgan,cheng2022autoregressive,liu2018learning} use voxel representation, \cite{Li2021}  use point clouds for learning latent spaces of shapes, and meshes are popular for the problem of 3D shape completion \cite{Gkioxari2019,shin2018pixels}.  Implicit representations \cite{Occupancy_Networks,michalkiewicz1901deep,Chen_2019_CVPR,Park_2019_CVPR} are gaining popularity for their lower footprint and greater representation power, however their application in the medical and biological domain has been minimal. 

In the medical and biological domain, raster representations have been prevailing even though it is identified that these are sub-optimal \cite{Yang_2022_CVPR}. The 3D adaptation \cite{CABR16} of U-net\cite{ronneberger2015u} was pioneering work for neural network-based methods for 3D data. This was followed by the field of medical imaging to be restricted to the problem of image segmentation and raster representations \cite{Milletari2016VNetFC,chen2021transunet,Alom2019,Zhou2020UNetRS}. \cite{Conrad2022} show that averaging the results of 2D segmentation through the third dimension performs better than applying 3D segmentation to 3D voxels of electron Microscopy data.  Even when using morphological priors for semi-automated segmentation of neuronal networks \cite{chen2021weakly}, raster representation was used for morphology encoding. The raster format has been posing challenges towards taking a volumetric approach towards cell organelles, as illustrated in  Fig. \ref{fig:teaser2} and the push towards implicit representations is fairly recent \cite{Yang_2022_CVPR,raju2022deep}.

Both \cite{Yang_2022_CVPR,raju2022deep} pertain to organ-level biological systems for which high-quality 3D imaging is possible through commercial clinical technologies such as computational tomography \cite{raju2022deep} and magnetic resonance imaging. Imaging nanoscale-sized organelles inside cells in 3D is more challenging, with the resolution limit of optical microscopes ($\sim200$ nm which is comparable to mitochondria diameters) and scarcely available and low-throughput 3D electron microscopes (EM). The shape modeling solutions for this scale are therefore scarce in the conventional voxel-representations and absent for implicit representations.

\paragraph{Nanoscale Scientific Shape Simulators:}
The use of simulators is not uncommon for cell microscopy \cite{SLEPCHENKO2003570, Heydari2017, Sekh2021} and \textit{molecular chemistry} \cite{Gupta2020, Gupta2020a, thornburg2022fundamental}. \cite{Heydari2017} model cell dynamics and \cite{Sekh2021} model microscope observations of mitochondria and vesicles. \cite{Sekh2021} uses a physics-based simulator to generate 2D ground truth annotations for the 3D shape of mitochondria in optical microscope images for simulation-supervised deep learning. Although this approach allows the creation of otherwise hard-to-obtain training data for supervised segmentation, the geometric shape of mitochondria is approximated using parametric functions. Another interesting application of simulators in learning is noted in \cite{Gupta2020a} and \cite{Gupta2020} that employ an adversarial learning paradigm with a physics simulator to learn 3D density maps for cryo-electron microscope (cryo-EM) reconstructions. It learns the shape of a specific molecule species through tens of thousands of noisy EM images of individual molecules of the same species in different orientations. Contrary to our problem, here, there is only one 3D shape to learn, and the variability in the data arises from different 2D projections and noise. While these works serve as inspiration, learning the shape of mitochondria is more complex due to large variability in shapes, scarcity of 3D EM datasets of mitochondria, and the need of deriving 3D segmentations from 3D EM datasets before developing shape models. 

\paragraph{Mitochondria Representation and Analysis:} 
A comprehensive and current list of mitochondria analysis tools for fluorescence microscopy images is provided in \cite{chu2022image}, which indicates the significance of mitochondria analysis. 
Mitochondria are analyzed today using conventional image processing \cite{lefebvre2021automated} and/or shape-primitives \cite{viana2015quantifying}.  Modeling mitcohondria as tubular geometries is a common simplification \cite{Sekh2021,viana2015quantifying,zamponi2018mitochondrial}. Therefore, thresholding, morphological operations and graph representation are employed. These provide insight into networks of mitochondria, but not the nanoscale morphology which is where the trigger for large scale behavioural changes lies.  Sophisticated user-friendly tools have been developed for mitochondria analysis \cite{viana2015quantifying, lefebvre2021automated, lihavainen2012mytoe}, but all with backbone of advanced image processing on rasterized images. In fact, \cite{lefebvre2021automated} resorts to pixel-to-pixel correspondences as the preferred non-parametric approach, the trade-off being that the analysis is precise up to pixel size which compares with the mitochondria size. Only recently, deep learning solutions have been proposed \cite{fischer2020mitosegnet, Sekh2021}, but still using raster representations. It was identified that generating ground truth in fluorescence microscopy images is quite difficult due to information loss in resolution, subjectivity of manual annotation, and tediousness.  Simulation-supervised dataset based on parametric representation was employed in \cite{Sekh2021} as a mechanism to eliminate the ground truth unavailability problem.  


\paragraph{Summary of Gaps and the Proposed Shift in Paradigm:} Raster and parametric representations of mitochondria are the conventionally used representation of mitochondrial morphology. Both suffer from the inability to model fine details of mitochondria shapes. EM datasets and annotation tools have emerged, but remain tied to raster representations. For computer vision approaches to play a larger role, simulation of sophisticated, diverse, and rich fluorescence microscopy image datasets with unambiguous ground truth is essential. We fill the gap by presenting solutions to derive implicit representations of mitochondria using EM dataset and use the implicit representations for a variety of computer vision tasks.



\section{Methodology}



\paragraph{Implicit Shape Representation:} \label{sub_sec:implicit}
We use occupancy network \cite{Occupancy_Networks} for shape representation of mitochondria. %
Occupancy is an implicit 3D shape representation that represents 3D surfaces as the continuous decision boundary of a deep neural network classifier. For a shape $S$, occupancy function $o$ maps every possible point in the 3D space $\mathbb{R}^3$ as the occupancy probability between 0 and 1 which can be denoted by, 
\begin{equation}
     o: \mathbb{R}^3 \rightarrow \{0,1\}
    \label{eq:occupancy}
\end{equation}
A deep neural network can be used to approximate this function denoted by $f_\theta$, which takes an observation $x \in \chi$ as input and outputs the occupancy for every point in $\mathbb{R}^3$.
\begin{equation}
    f_\theta : \chi \times \mathbb{R}^3 \rightarrow [0,1]
    \label{eq:occ_gen1}
\end{equation}
This network is termed as occupancy network. Occupancy-based implicit representation is especially suited for our purpose since it allows us to shift to isomorphic representations from the voxelized segmentations of EM images. 


\subsection{Generative Shape Model}\label{sub_sec:generative}
\begin{figure}
    \centering
    \includegraphics[width=0.8\linewidth]{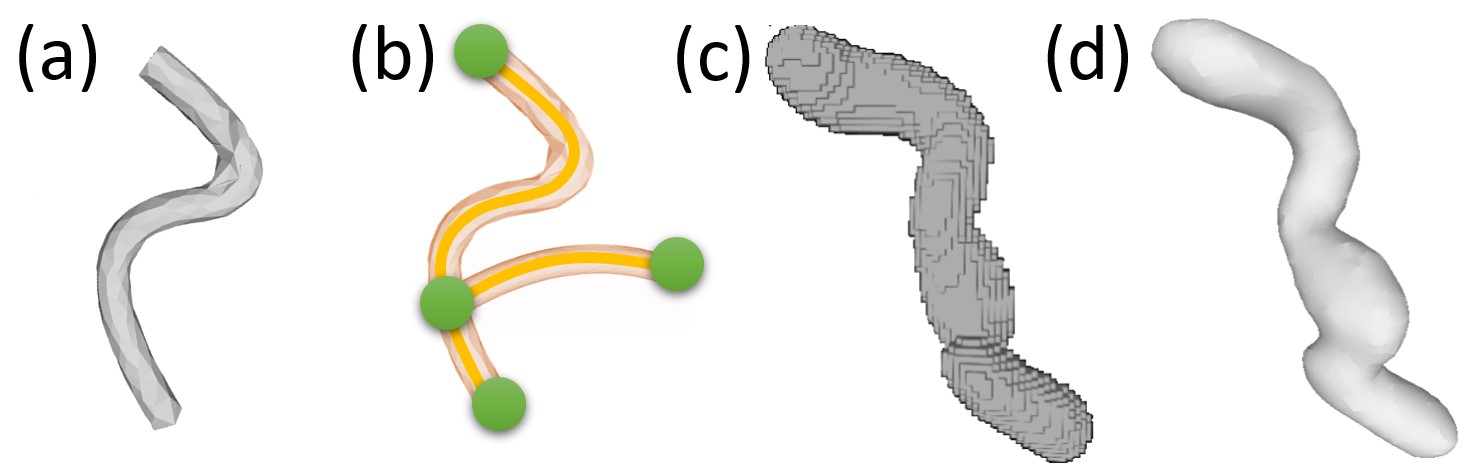}
    \vspace{-3mm}
    \caption{Tubular parametric shapes are used in the state-of-the-art (a) \cite{Sekh2021} and (b) \cite{viana2015quantifying}. They are ineffective at representing the complexity and shape diversity in mitochondria. Isotropic high-resolution 3D EM images (c) are better at this. Meshes (d) extracted through the EM images represent the shapes of mitochondria better. }
    \label{fig:paramteric_not_enough}
\end{figure}

The complexity of the shape of mitochondria is lost in parametric representations, but it can be learnt as a prior on the shape distribution of mitochondria. Consider the examples shown in Fig. \ref{fig:paramteric_not_enough}. The two parametric representations are observed to be ineffective at representing the small details and complexity of shapes of mitochondria. Isotropic high-resolution 3D EM images are better at retaining these details and presenting the structure of mitochondria. Therefore, we use them to derive the implicit shape models.

\textbf{Data Pre-processing:} The starting point of our approach is the 3D segmentations generated from 3D EM microscopy datasets. Individual mitochondria are extracted from the 3D segmentations using a 3D connected components analyzer \cite{william_silversmith_2021_5535251}. Meshes of individual mitochondria are extracted using the marching cubes algorithm. 
Then the meshes are prepared for the occupancy networks as follows. First, occupancy network requires 
that the meshes are watertight (hole-free) to conceptualize a point as existing either inside or outside a mesh. So, any holes arising due to missing data are fixed to maintain the gradient of the broken mesh.  
Each mesh is then normalized to fit into a unit cube of dimension 1. No stretching or skew is introduced and the operation is equivalent to performing isotropic scaling of Cartesian coordinates. 
For a dataset of 3D segmentations derived from a given electron microscope, the range of scales used may be stored for future reference. 
Lastly, 10000 points are samples uniformly in the unit cube and their occupancy value are computed. This data pair of query points and occupancy values forms the occupancy representation and is used as the training data.

\textbf{The Unconditional Generative Model:} Our generative model for representing mitochondria shapes uses the basic mode of occupancy network without any conditioning
\begin{equation}
    f_\theta(h, p) = o: \mathbb{R}^c \times \mathbb{R}^3 \rightarrow \mathbb{R}
    \label{eq:occ_gen2}
\end{equation}
where $h$ is a $c$- dimensional latent vector encoding of the shape $S$ and $ p \in \mathbb{R}^3  $ is the query point and $ o \in [0,1]$, is the probability of occupancy. 

We use the same methodology as occupancy networks for our unconditional generative model. The sampled points are used as input to the model and the model is trained to predict the occupancy values of the query point. To learn a latent distribution or shape prior, the model is constructed as a variational auto encoder (VAE). 
The loss is a combination of classification cross-entropy loss between the true occupancies and the predicted occupancies and the KL divergence between the latent space and a Gaussian distribution.  
Once the training is complete, the model can be used in the inference stage to randomly generate new shapes by passing Gaussian noise as input.  The architecture and implementation details are provided in the supplementary. 





\subsection{Fluorescence Images from 3D EM Images} \label{sub_sec:EM2Fl}

Creating fluorescence microscopy images or z-stacks from 3D EM data comprises of three major steps, namely generating the shape from the EM image, simulating the fluorescence labeling by computing fluorescence molecule distribution on the shape and generating the fluorescence image from the fluorescence molecule distribution. These are described in the following paragraph.



\textbf{Individual Mitochondria Shape Extraction:} The pre-processing steps described in section \ref{sub_sec:generative} are applied on the segmented 3D EM image to extract meshes of the individual instances of mitochondria from EM volume. 

\textbf{Fluorescence Molecule Distribution:} Having generated the mitochondria meshes, we scale the meshes back to physical dimensions. We simulate fluorescence labeling by sampling points on the shape and using them as the fluorescence molecules' binding locations. The mesh size is determined by the density of fluorescence molecules that the mitochondrion is expected to be labeled with. For example, if an outer membrane fluorescent label is used and the expected surface density of fluorescent molecules is 30 molecules$/\upmu$m$^2$, then a surface mesh is created such that there are $\sim$30 nodes$/\upmu$m$^2$. This list of Cartesian coordinates of the fluorescent molecules is the input for the next stage. 

The mesh is rotated as necessary for the desired perspective. For this, we have taken the origin to be generally at the center of the  fluorescent molecules co-ordinates, but in general, it can be chosen arbitrarily. Rotation is achieved as follows. Given the original Cartesian coordinates of the fluorescence molecules denoted by $(x',y',z')$, then the new coordinates $(x,y,z)$ are obtained as 
\begin{equation}
    (x,y,z)^{\rm T}= R_z(\gamma)R_y(\beta)R_x(\alpha) (x',y',z')^{\rm T}
\end{equation}
where $\alpha, \beta, \gamma$ are the Euler's angles about the $x,y,z$ axes. 

\textbf{Generating Fluorescence Microscopy Image or Z-stack:} Every fluorescence microscope is characterized by a point spread function (PSF). Computationally, the microscope image is a noisy version of the PSF convolved over the locations of fluorescence molecules. The PSF of a microscope is specified by its optical design, the numerical aperture of the microscope objective lens, the refractive index of the immersion medium between the objective lens and the sample, the geometry of the sample holder, and the wavelength of the fluorescence signal being emitted by the fluorescence molecules. 
We refer the microscope simulators \cite{sekh2021_codes} publicly released by the authors of Sekh et. al\cite{Sekh2021}. Noise in the images is simulated using the noise model presented in the same source.

\begin{figure}[t]
    \centering
    \includegraphics[width=0.9\linewidth]{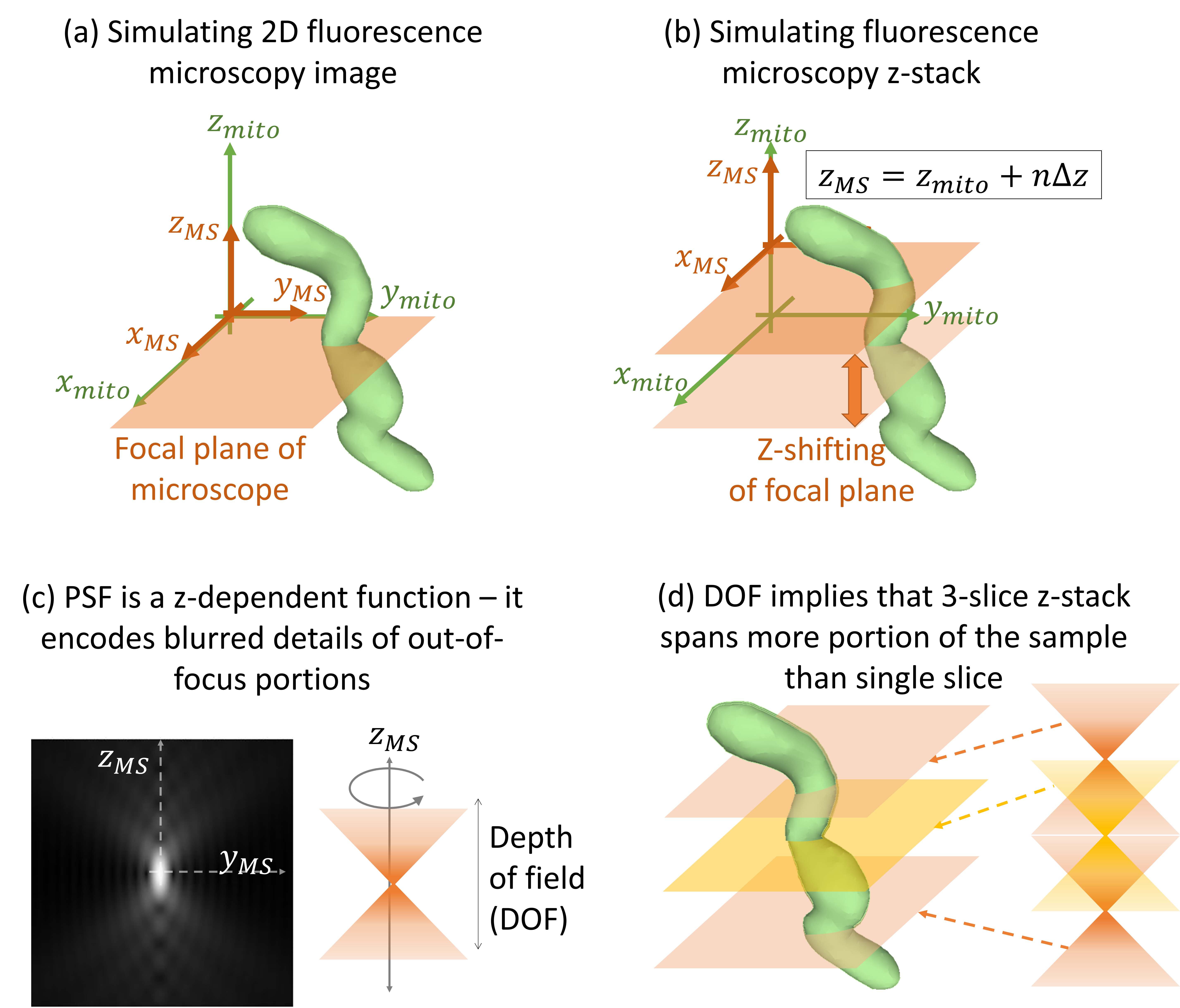}
    \caption{Coordinate systems for generating fluorescence microscopy image and z-stack, and illustration of the point spread function of an epifluorescence microscope.}
    \label{fig:z-stack}
\end{figure}

\cite{Sekh2021} simulate only a single 2D fluorescence image of the sample corresponding to the focal plane of the microscope. Let the coordinate system of the mesh of the mitochondrion be $(x_{mito},y_{mito},z_{mito})$ and the coordinate system of the microscope be $(x_{ms},y_{ms},z_{ms})$. Plane $z_{ms}=0$ indicates the focal plane of the microscope. When forming the 2D image, $z_{mito}=0$ indicates the slice of the mitochondrion that is in focus. This is illustrated in Fig. \ref{fig:z-stack}a. However, if a z-stack is needed, the focal plane of the microscope is scanned with respect to the sample with a desired step size $\Delta z$. After introducing an offset $n\Delta z$ between the sample and microscope Cartesian coordinates where $n$ is an integer. For simulating a 3-slice z-stack, we use $n = -1, 0, 1$. This is illustrated in Fig. \ref{fig:z-stack}b.

In the supplementary, we present another exciting application of our shape representations, namely generating fluorescence images of mitochondria corresponding to a target microscope from fluorescence images or stacks of the same mitochondria acquired using another microscope. 

\subsection{3D Shape from 2D Fluorescence Image/Z-stack}\label{sub_sec:3D from 2D}
Generating 3D shapes from 2D images is not new in computer vision \cite{Occupancy_Networks, Chen_2019_CVPR, NIPS2019_8340, Wu_2020_CVPR, chen2019dibrender}, but is considered far-fetched in the microscopy domain. Here, we address this highly ill-conditioned problem of generating 3D shapes from a single 2D fluorescence microscopy image. We propose another better-conditioned solution where 3-slice z-stack is used instead of a single image to reconstruct the 3D shape. 
Conditioning individual mitochondria shapes with their counterpart fluorescence microscopic image or z-stack permit learning the shape correspondence between microscope images and the corresponding 3D shapes. This model can be written as, 
\begin{equation}
    f_\theta(h, p, m) = o: \mathbb{R}^M \times \mathbb{R}^c \times \mathbb{R}^3 \rightarrow \mathbb{R}
    \label{eq:occ_gen3}
\end{equation}
where $m$ is the encoding of the fluorescent microscope image or z-stack corresponding to the shape $S$. For this, we train an occupancy model conditioned on the encoding of simulated microscope images. We train two separate models, one for mapping a single 2D fluorescence image to 3D shape (image-to-shape) and the other for mapping 3-slice fluorescence z-stack to 3D shape (stack-to-shape). The implementation details are presented in the supplementary.

\textbf{3D Context Embedded in 2D Fluorescence Images:} Indeed the 2D fluorescence image to 3D shape reconstruction is heavily ill-conditioned problem. However, as shown in Fig. \ref{fig:z-stack}c, the PSF has a 3D nature, that is generally rotationally symmetric in the lateral x-y plane but presents z-dependent blurring to the sample. Therefore, even if we use a single 2D fluorescence image, if has the 3D context of the sample within the depth-of-field (DOF) of the microscope and  form of z-position encoding through the PSF. Consequently, the 2D image to 3D shape reconstruction problem is not as ill-conditioned as it appears at the first sight. Nonetheless, the 3D shape beyond the DOF does not get imaged and therefore the reconstructed shapes are only accurate within the DOF.

Further, the use of 3-slice z-stack presents a better conditioning than the use of a single 2D image. Here, we use 3-slices such that $\Delta z = $DOF$/2$, as shown in Fig. \ref{fig:z-stack}d. 
This selection contributes two advantages. First, it allows to span axially larger extent of the sample, supporting better shape reconstruction. Second, it allows the central region of the shape to be sampled twice, which improves the reconstruction quality of this portion.

In principle, more z-slices can be included to span larger samples, for example, if the target is to reconstruct an entire cell, which is 5-20 $\upmu$m thick in comparison to DOF of $\sim500$ nm for a high-resolution fluorescence microscope.


\section{Results}

\paragraph{Dataset:}
We use the segmentation of the publicly available 3D EM data from \cite{Parlakgul2022} to generate the data required for training our models. The data is obtained using a Focused Ion Beam Scanning Electron Microscopy (FIB-SEM) which provides images of sub-cellular structures with an isotropic resolution of $8$nm. The segmentations of mitochondria provided with this data are obtained using a semi-automated deep learning-based approach, where a few samples are annotated for training. The model is then fine-tuned to generate segmentation masks for the entire data volume. The masks are verified quantitatively \cite{Parlakgul2022}. The segmented EM is re-sampled at 24 $nm$ for computational efficiency.




\begin{figure}[t]
  \centering
   \includegraphics[width=0.9\linewidth]{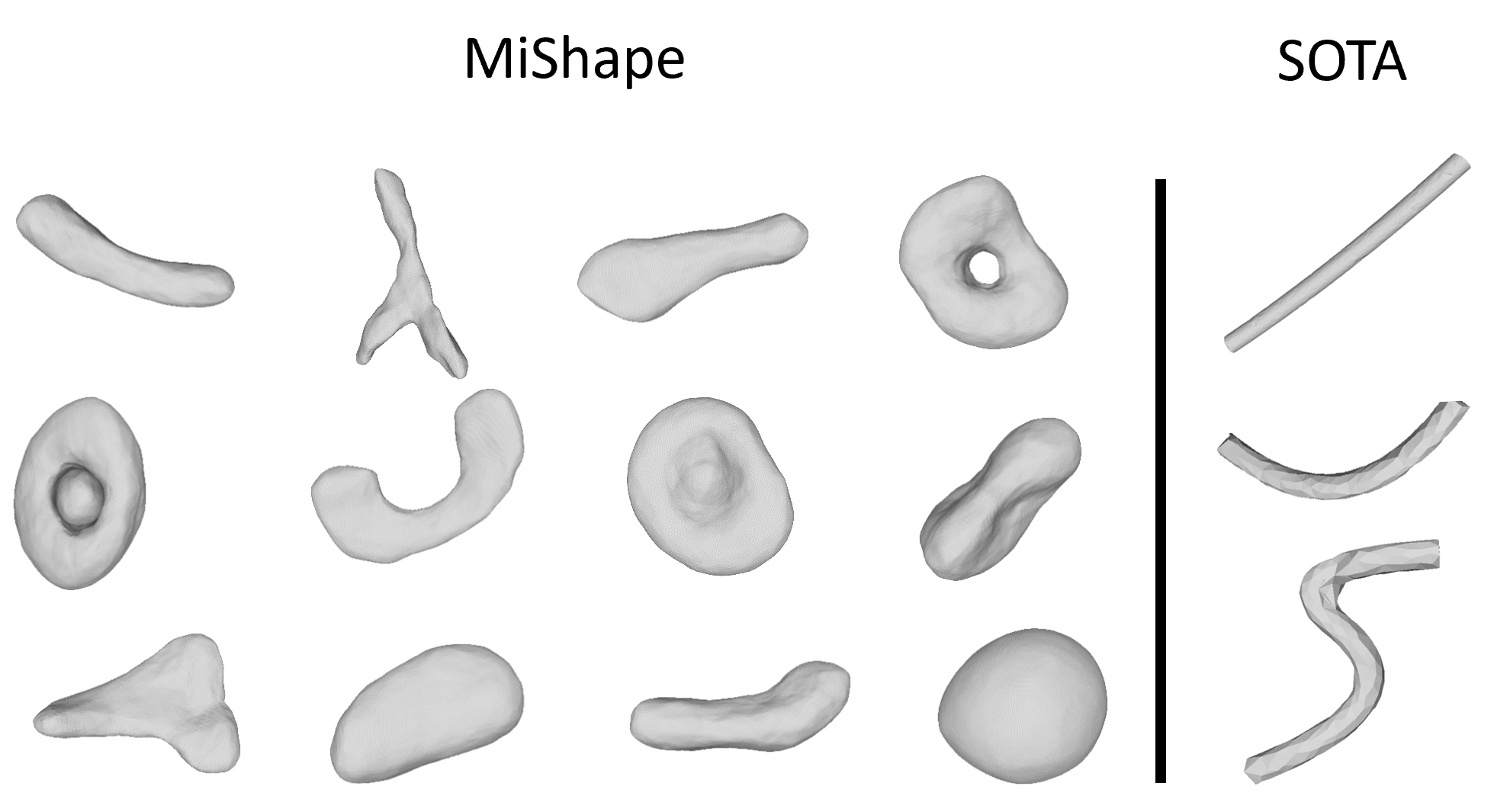}
   \caption{Generated 3D shape samples of mitochondria using MiShape present greater diversity and complexity than the state-of-the-art \cite{Sekh2021}.}
   \label{fig:gen_res}
\end{figure}

\subsection{Unconditional Generative Model}\label{sub_sec:res_generative}
We train the unconditional generative shape model discussed in section \ref{sub_sec:generative} on mitochondrial shapes acquired from extracting 3D EM data. The generative model is trained on 27272 instances of 3D mitochondria shapes represented using implicit representation. In Fig. \ref{fig:gen_res}, we display randomly selected generated samples of the unconditional generative shape prior, MiShape. It is seen that the shapes generated by our method are more complex and varied than the current state-of-the-art \cite{Sekh2021} parametric representation which assumes mitochondria as tubular structures. We also verified that our structures match the structures reported in \cite{miyazono2018uncoupled}. It demonstrates that our generator is effective in modeling and generating realistic mitochondrial shapes.

\begin{figure}
    \centering
    \includegraphics[width=1\linewidth]{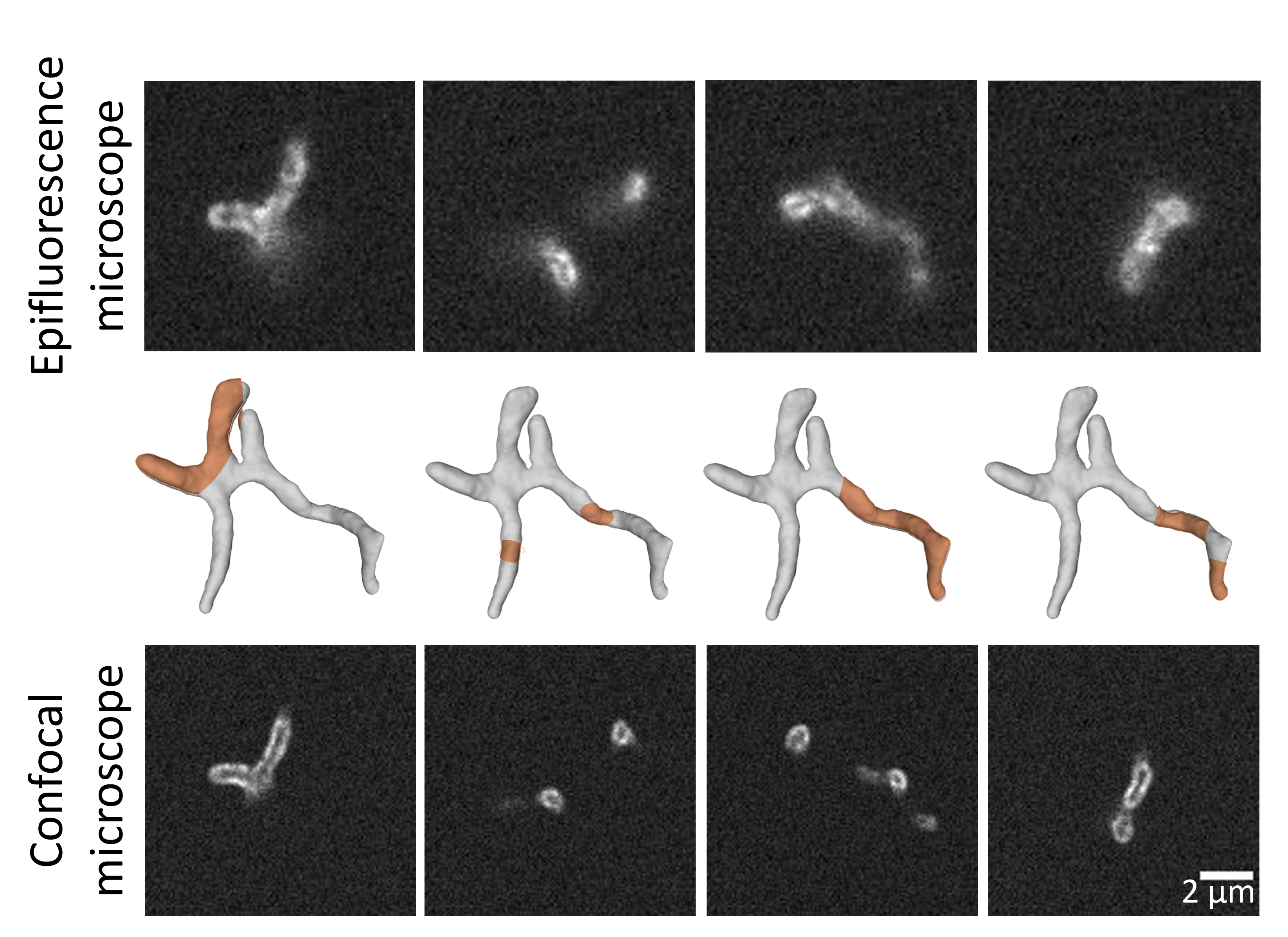}
    \caption{Illustrations of fluorescence microscopy images generated from different rotations of a mitochondrial network derived from 3D EM image segmentation (EM data in supplementary). The regions in orange on the 3D shape denote the regions which fall in the DOF of the focal plane.}
    \label{fig:res_Ep-to-F}
\end{figure}


\subsection{EM Image to Fluorescence Microscopy Images}\label{sub_sec:res_Ep-to-Fl}
We present in Fig. \ref{fig:res_Ep-to-F} examples of fluorescence microscopy images created from a complex mitochondrial network shape present in a 3D EM image. The 3D EM image segmentation, after being converted to meshes, have been used to simulate images of different perspectives for two fluorescence microscopes. We specifically use two different types of fluorescence microscopes, one being epifluorescence microscope (Epi1) and the other being confocal microscope (Con1) The exact parameters used for Epi1 and Con1 are described in the supplementary. They have different optical designs, different digital resolutions, as well as different image qualities. Epifluorescence microscopes are high throughput microscopes but with poor z-sectioning (DOF $\sim500$ nm) and poor lateral resolution ($\sim250$ nm) while confocal microscopes are low throughput but have superior z-sectioning (DOF $\sim 250$ nm) and superior lateral resolution ($\sim180$ nm). The digital resolutions of Epi1 and Con1 are 109 nm and 80 nm respectively. It is seen that implicit shapes allow simulation of diverse 2D fluorescence microscopy images at different resolutions, perspectives, and microscopes using a single 3D segmentation. Through this process, we generated large training datasets for the experiments in sections \ref{sub_sec:res_2D_to_3D}, \ref{sub_sec:res_2Dsegmenation} and the microscope-to-microscope transformation in the supplementary.

\subsection{Shapes from Fluorescence Images and Stacks}\label{sub_sec:res_2D_to_3D}

\textbf{Epifluorescence Stack-to-Shape:} 
For each of the 27272 mitochondria instances in the 3D EM dataset, we simulated epifluorescence (Epi1) z-stacks for 6 perspectives (i.e. 6 combinations of the Euler's angles $\alpha,\beta, \gamma$). Each stack comprised of three slices specified by $z_{ms}=-250, 0, 250\,nm$. We use a 3D encoder to encode the conditional stack input. 
After training the conditional generative model, the stacks in test set were used for inference. 

\textbf{Confocal Image-to-Shape:} Here, we present a challenging situation to our approach. We use a single image to reconstruct the 3D shapes and enforce that these images correspond to confocal microscope Con1, which has a smaller depth of field than epifluorescence microscope. Therefore, the presence of 3D context in the fluorescence image is quite limited. 
The procedure is the same as described before, except that single 2D confocal microscopy images are used instead of epifluorescence stacks and a 2d encoder is used for encoding the conditioning image input.


\begin{table}
    \centering
    \scalebox{0.9}{
    \begin{tabular}  {@{}l c@{} p{3mm}  c@{}}
     \toprule
       Pre-train Data      & IoU   &     & Chamfer L1 \\
     \midrule
      Epifluorescence stack to shape  & 0.46  & & 0.078 \\
      Epifluorescence image to shape  & 0.44  & & 0.083 \\
      Confocal image to shape            & 0.44 &   & 0.087\\
     \bottomrule
       
    \end{tabular}}
    \caption{Quantitative comparison of the performance of 3D shape reconstruction using stack and images.}
       \label{tab:shape_from_fluorescence}
   
  \end{table}

 \begin{figure}
       \centering
    \includegraphics[width=1\linewidth]{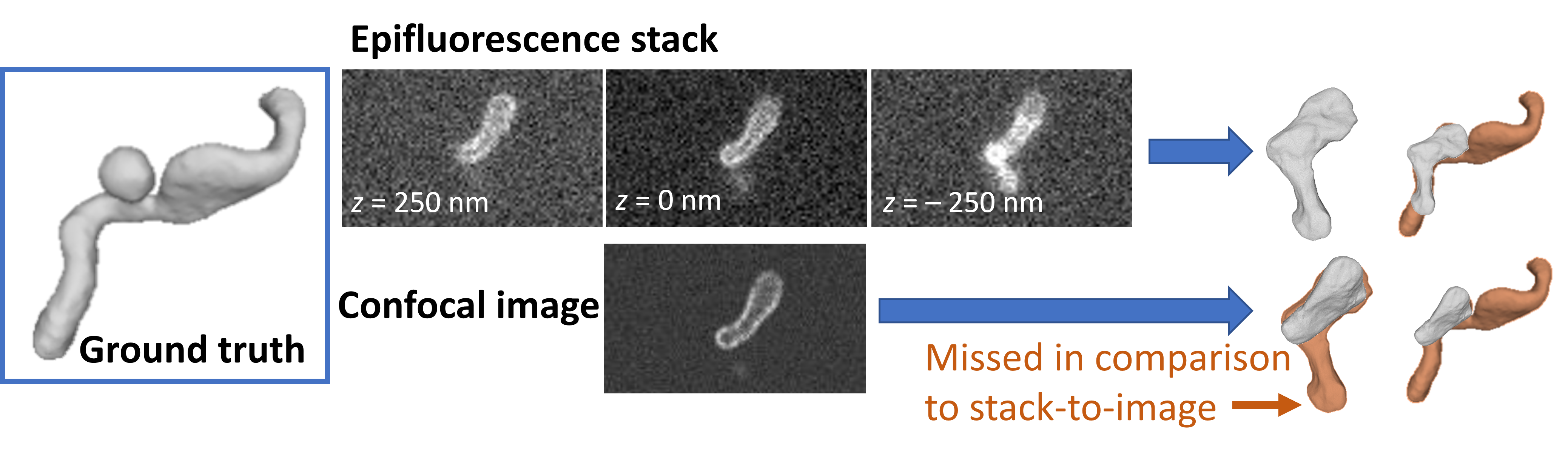}
       
       \caption{Comparative illustration of the reconstruction performance of epifluorescence stack-to-shape model (top-row) and confocal image-to-shape model (bottom row). Orange color depicts the features missed in comparison to a reference.}
       \label{fig:CompareFl2EM}
   \end{figure}

\textbf{Reconstruction Results:} The reconstruction accuracy of the stack-to-shape and image-to-shape models are presented in Table \ref{tab:shape_from_fluorescence}. We follow \cite{Occupancy_Networks} to compare our different conditional modes. 
%

It is observed that the epifluorescence stack-to-shape model improves volumetric IoU (46\% IoU compared to 44\% in image-to-stack) despite using only 3-slices
This indicates further improvement in reconstruction when using more slices. 
The image-to-shape models perform better in terms of Chamfer L1 metric for both epifluorescence and confocal microscope as the region generated is smaller and thus the error of margin is lower for correspondences for Chamfer L1. This is illustrated in Fig. \ref{fig:CompareFl2EM}, where we see that the stack-to-image model reconstructs only part of the shape and loses fine details such as the blob in the top-left of the ground truth.
In order to present insight into how the smaller DOF of the confocal microscope contributes to this, we consider an example confocal image which has a one-to-one correspondence with the middle slice of the epifluorescence stack shown in Fig. \ref{fig:CompareFl2EM}. The shape reconstructed using the confocal image spans a further smaller portion than the portion reconstructed by the stack-to-shape model. 

\begin{figure}
  \centering
    \includegraphics[width=1\linewidth]{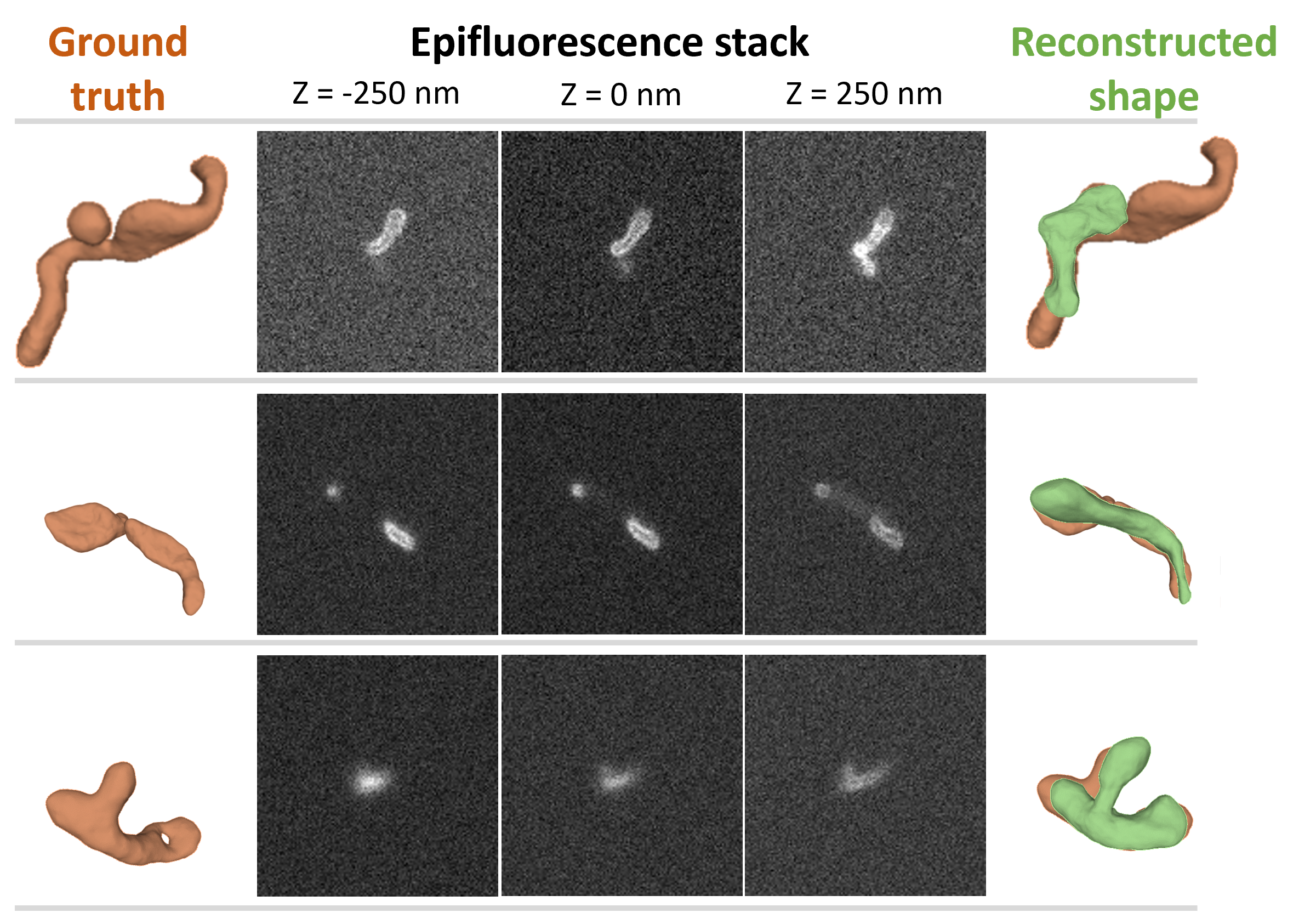}
    
  \caption{Examples of 3D Shape generated given a 3 slices of epifluorescence z-stack 2D.} 
  \label{fig:Epistack_to_3d}
\end{figure}

\begin{figure}
  \centering
  \begin{subfigure}{\linewidth}
  \centering
    \includegraphics[width=\linewidth]{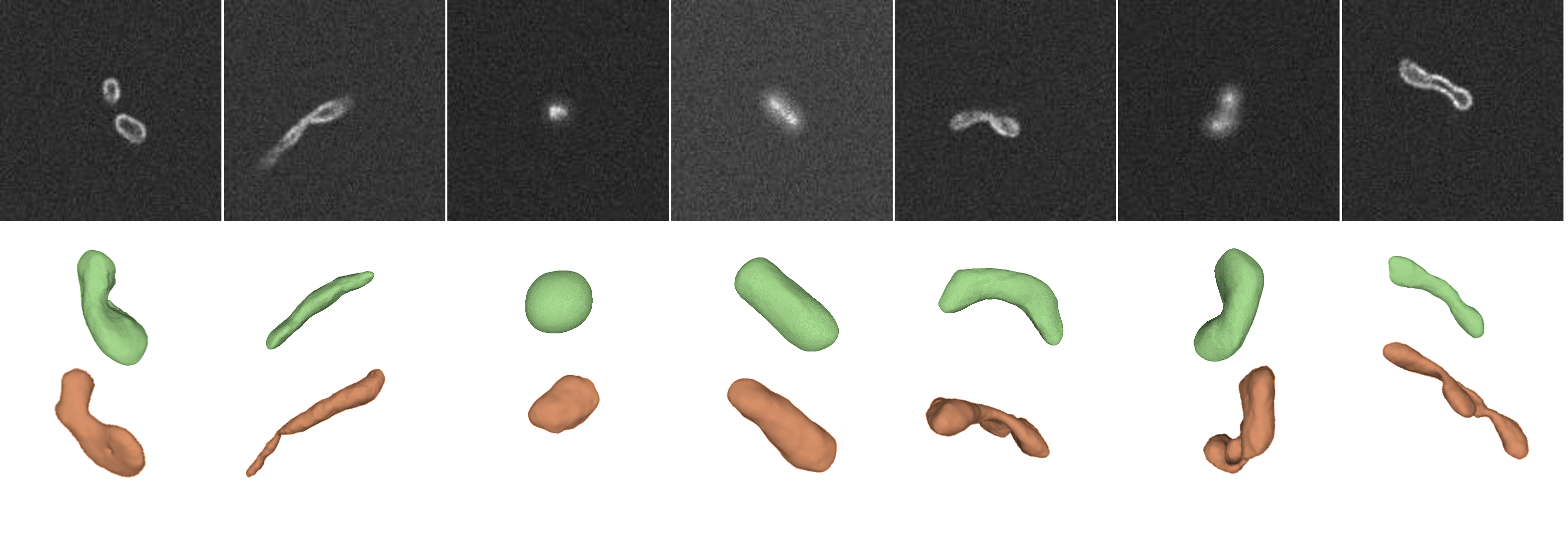}
  \end{subfigure}
  \caption{Examples of 3D Shape generated given a single 2D confocal microscope image.}
  \label{fig:Confocal2d_to_3d}
\end{figure}

Qualitative results for the above experiments are presented in Fig. \ref{fig:Epistack_to_3d} and Fig. \ref{fig:Confocal2d_to_3d}. Ground truth shapes are shown in orange. It is observed that the 3D shapes reconstructed using our stack-to-shape model are close to the ground truth shapes where the z-span of the stack or the DOF of the microscope allows. We present results for epifluorescence image-to-shape in the supplementary.


\subsection{2D Segmentation of Mitochondria from Fluorescence Microscopy Images}\label{sub_sec:res_2Dsegmenation}
\begin{figure}
  \centering
    \includegraphics[width=1\linewidth]{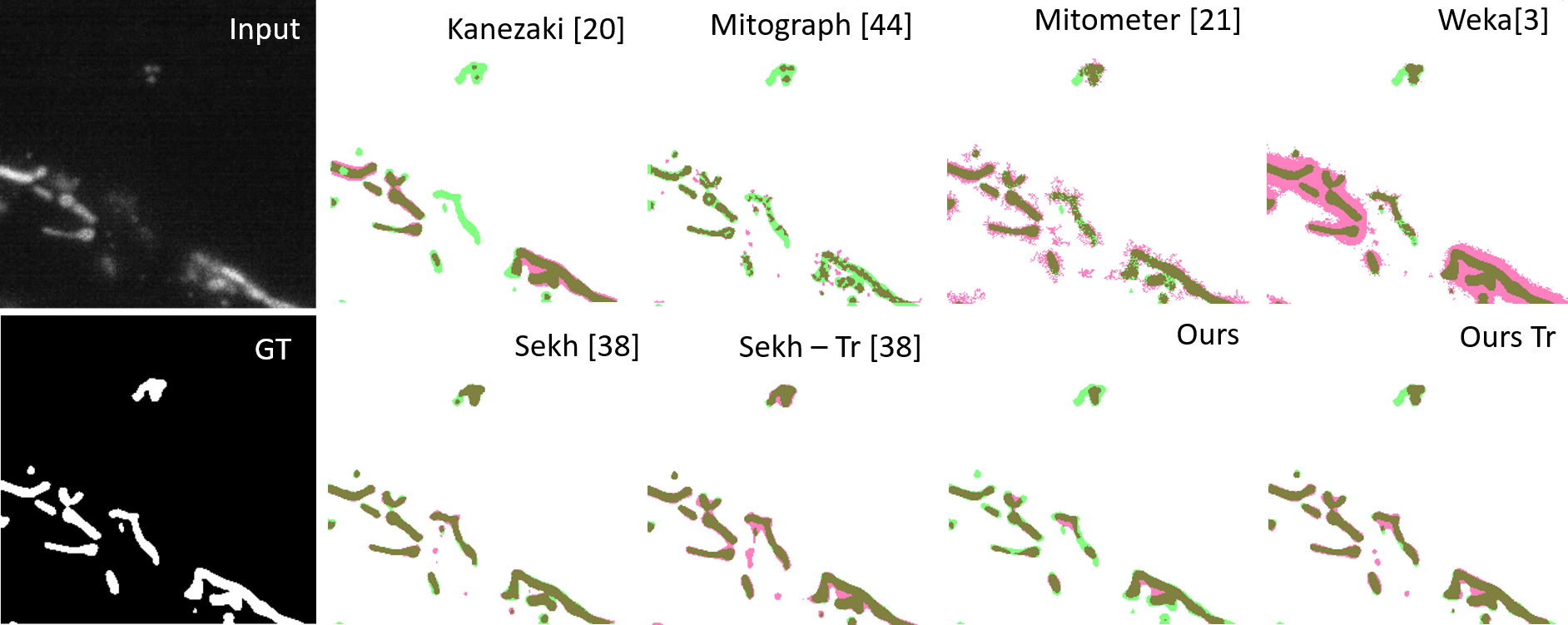}
  \caption{Example 3D shapes generated using 3-slice of epifluorescence z-stack. Green regions in the segmentations are false negative (missed pixels), red regions are false positives, other white or dark regions are correct detections.} 
  \label{fig:seg_res}
\end{figure}

In our final experiment, we create a dataset of optical microscope images and their 2D ground truth segmentation masks for the problem of 2D segmentation. 
Shapes extracted from EM is more representative than shape approximations using parametric curves as done by \cite{Sekh2021}. 
We create a simulated dataset of 7000 images matching the parameters of the real microscope data from an epifluorescence microscope \cite{Sekh2021}. To test on real data, we use the manually annotated dataset from \cite{Sekh2021} consisting of 279 training images and 144 test images. We run inference using (i) a model trained using only simulated data and, (ii) a model trained on simulated data and fine-tuned using the training split of manually annotated images. The segmentation model achieves a dice score of 0.77 and F1 score of 0.64 when trained using only simulated dataset. Fine-tuning the model on the manually annotated training set achieves a dice score of 0.86 and F1 score of 0.76. Qualitative results of the segmentation are provided in Fig \ref{fig:seg_res}. We include quantitative benchmarks of commonly 2D segmentation tools for mitochondria in optical microscope images \cite{viana2015quantifying, lefebvre2021automated, 10.1093/bioinformatics/btx180, 10.1109/ICASSP.2018.8462533, Sekh2021}. 



\section{Discussion}

\textbf{Limitations using EM Data:} 
The biggest advantage of 3D electron microscopy is its resolution.
However, the versatile use of 3D EM datasets for learning sub-cellular shape priors is difficult due to several limitations. The sample preparation for FIB-SEM images is quite challenging, especially if the morphology of the cells and sub-cellular organelles is to be preserved to match their living state. Second, the lack of specificity in EM implies that everything in the cell gets imaged. Correlative light electron microscopy supplemented with modern processing helps, but requires sophisticated experimental protocols. The third limiting factor is the throughput of 3D EM imaging. Imaging single cells require several hours, sometimes days of imaging. Public datasets are emerging, yet the availability of 3D EM images and high-quality segmentations is still scarce. 

\textbf{Possibility of using 3D Fluorescence Microscopy Data for constructing Shape Priors:} Fluorescence microscopy provides relatively simpler experimental protocols, compatibility with living cell imaging, specificity,  simpler and faster mechanisms of 3D imaging, and relatively higher throughput. Therefore, it is exciting to consider 3D fluorescence microscopy images or z-stacks as a candidate for learning shape prior. 
The main limitation of fluorescence microscopy is resolution and its anisotropicity (the z- resolution is different from xy-resolution). Confocal and Airy scan microscopes permit isotropic sampling of 3D space and support a smaller resolution ($\sim$ 180 nm) than most microscopes, but the details in mitochondria are even smaller. 

A likely solution is to use nanoscopes, i.e. super-resolution fluorescence microscopes. Stimulated Emission Depletion (STED) microscopy provides resolution as small as 20-30 nm in 3D, albeit with a compromise in throughput. Alternatively, higher throughput and almost isotropic resolution can be achieved using Structured Illumination Microscopy (SIM), although with poor resolution ($\sim 120$ nm). Notably, nanoscopy of mitochondria images is a recent development \cite{stephan2019live,jakobs2020light,liu2022multi,pape2020multicolor,samanta2019fluorescent}. Nonetheless, we look forward to the emergence of large public datasets of nanoscopy images and their use in learning shape priors of mitochondria and other sub-cellular organelles.    

\textbf{Future Directions:} We surmise that our work will open the path to developing an implicit shape atlas of cell organelles. This will allow for migration towards the simulation of realistic cell and sub-cellular morphologies. Also, this will facilitate the creation of sophisticated and realistic microscopy datasets with unambiguous ground truths, and enable computer vision and artificial intelligence to play a larger role in knowledge discovery in life sciences. We think that this work will also facilitate modeling the dynamic deformations and interactions of mitochondria and investigation of these aspects through fluorescence microscopy videos.


\section{Conclusion}

In this paper, we introduce a generative model-based shape prior, MiShape, that represents the wide range of shapes of mitochondria at high resolution.
Our experiments demonstrate the reconstruction of 3D mitochondria shapes from single 2D fluorescence microscopy images and stacks containing only a few z-slices.
We show the versatility of our method in constructing large and complex simulated fluorescence microscopy datasets for 3D reconstruction from fluorescence images, 2D segmentation, and microscope-to-microscope transformation (discussed in supplementary). 
We believe that this work will initiate the use of implicit shape priors for the hard problem of 3D reconstruction from fluorescence microscopy and open up the possibilities for solving a variety of problems in life sciences. 

MiShape helps the sub-cellular biological vision domain to migrate 
to flexible and versatile implicit representations.
We believe that this work will accelerate the progress toward modeling an atlas of the cell together with the numerous organelles it contains. 


\clearpage
{\small
\bibliographystyle{ieee_fullname}
\bibliography{egbib}

\begin{thebibliography}{10}\itemsep=-1pt

\bibitem{abdollahzadeh2021deepacson}
Ali Abdollahzadeh, Ilya Belevich, Eija Jokitalo, Alejandra Sierra, and Jussi
  Tohka.
\newblock Deepacson automated segmentation of white matter in 3d electron
  microscopy.
\newblock {\em Communications Biology}, 4(1):1--14, 2021.

\bibitem{Alom2019}
Md~Zahangir Alom, Chris Yakopcic, Mahmudul Hasan, Tarek~M. Taha, and Vijayan~K.
  Asari.
\newblock {Recurrent residual U-Net for medical image segmentation}.
\newblock {\em https://doi.org/10.1117/1.JMI.6.1.014006}, 6(1):014006, mar
  2019.

\bibitem{10.1093/bioinformatics/btx180}
Ignacio Arganda-Carreras, Verena Kaynig, Curtis Rueden, Kevin~W Eliceiri,
  Johannes Schindelin, Albert Cardona, and H Sebastian~Seung.
\newblock {Trainable Weka Segmentation: a machine learning tool for microscopy
  pixel classification}.
\newblock {\em Bioinformatics}, 33(15):2424--2426, 03 2017.

\bibitem{chen2021transunet}
Jieneng Chen, Yongyi Lu, Qihang Yu, Xiangde Luo, Ehsan Adeli, Yan Wang, Le Lu,
  Alan~L. Yuille, and Yuyin Zhou.
\newblock Transunet: Transformers make strong encoders for medical image
  segmentation.
\newblock {\em CoRR}, abs/2102.04306, 2021.

\bibitem{chen2019dibrender}
Wenzheng Chen, Jun Gao, Huan Ling, Edward Smith, Jaakko Lehtinen, Alec
  Jacobson, and Sanja Fidler.
\newblock Learning to predict 3d objects with an interpolation-based
  differentiable renderer.
\newblock In {\em Advances In Neural Information Processing Systems}, 2019.

\bibitem{chen2021weakly}
Xuejin Chen, Chi Zhang, Jie Zhao, Zhiwei Xiong, Zheng-Jun Zha, and Feng Wu.
\newblock Weakly supervised neuron reconstruction from optical microscopy
  images with morphological priors.
\newblock {\em IEEE Transactions on Medical Imaging}, 40(11):3205--3216, 2021.

\bibitem{Chen_2019_CVPR}
Zhiqin Chen and Hao Zhang.
\newblock Learning implicit fields for generative shape modeling.
\newblock In {\em IEEE Conference on Computer Vision and Pattern Recognition
  (CVPR)}, June 2019.

\bibitem{cheng2022autoregressive}
An-Chieh Cheng, Xueting Li, Sifei Liu, Min Sun, and Ming-Hsuan Yang.
\newblock Autoregressive 3d shape generation via canonical mapping.
\newblock In {\em European Conference on Computer Vision}, 2022.

\bibitem{chinnery2021primary}
Patrick~F Chinnery.
\newblock Primary mitochondrial disorders overview.
\newblock {\em GeneReviews{\textregistered}[Internet]}, 2021.

\bibitem{chu2022image}
Ching-Hsiang Chu, Wen-Wei Tseng, Chan-Min Hsu, and An-Chi Wei.
\newblock Image analysis of the mitochondrial network morphology with
  applications in cancer research.
\newblock {\em Frontiers in Physics}, page 289, 2022.

\bibitem{CABR16}
{\"O}. {\c{C}}i{\c{c}}ek, A. Abdulkadir, S.S. Lienkamp, T. Brox, and O.
  Ronneberger.
\newblock 3d u-net: Learning dense volumetric segmentation from sparse
  annotation.
\newblock In S. Ourselin, W.S. Wells, M.R. Sabuncu, G. Unal, and L. Joskowicz,
  editors, {\em International Conference on Medical Image Computing and
  Computer-Assisted Intervention (MICCAI)}, volume 9901 of {\em LNCS}, pages
  424--432. Springer, Oct 2016.
\newblock (available on arXiv:1606.06650 [cs.CV]).

\bibitem{Conrad2022}
Ryan Conrad and Kedar Narayan.
\newblock {Instance segmentation of mitochondria in electron microscopy images
  with a generalist deep learning model}.
\newblock {\em bioRxiv}, page 2022.03.17.484806, mar 2022.

\bibitem{Feichtenhofer_2019_ICCV}
Christoph Feichtenhofer, Haoqi Fan, Jitendra Malik, and Kaiming He.
\newblock Slowfast networks for video recognition.
\newblock In {\em Proceedings of the IEEE/CVF International Conference on
  Computer Vision (ICCV)}, October 2019.

\bibitem{mitosegnet_code}
Christian~A Fischer, Laura Besora-Casals, St{\'e}phane~G Rolland, Simon
  Haeussler, Kritarth Singh, Michael Duchen, Barbara Conradt, and Carsten Marr.
\newblock Mitos segmentation tool (gpu) for linux.
\newblock \url{https://zenodo.org/record/3556431}, 2019.
\newblock Online; accessed 18-November-2022.

\bibitem{mitosegnet_model}
Christian~A Fischer, Laura Besora-Casals, St{\'e}phane~G Rolland, Simon
  Haeussler, Kritarth Singh, Michael Duchen, Barbara Conradt, and Carsten Marr.
\newblock Mitosegnet segmentation model.
\newblock \url{https://zenodo.org/record/3539340##.Xd-oN9V7lPY}, 2019.
\newblock Online; accessed 18-November-2022.

\bibitem{fischer2020mitosegnet}
Christian~A Fischer, Laura Besora-Casals, St{\'e}phane~G Rolland, Simon
  Haeussler, Kritarth Singh, Michael Duchen, Barbara Conradt, and Carsten Marr.
\newblock Mitosegnet: easy-to-use deep learning segmentation for analyzing
  mitochondrial morphology.
\newblock {\em Iscience}, 23(10):101601, 2020.

\bibitem{frangi1998multiscale}
Alejandro~F Frangi, Wiro~J Niessen, Koen~L Vincken, and Max~A Viergever.
\newblock Multiscale vessel enhancement filtering.
\newblock In {\em International conference on medical image computing and
  computer-assisted intervention}, pages 130--137. Springer, 1998.

\bibitem{Gkioxari2019}
Georgia Gkioxari, Justin Johnson, and Jitendra Malik.
\newblock {Mesh R-CNN}.
\newblock {\em IEEE International Conference on Computer Vision},
  2019-October:9784--9794, jun 2019.

\bibitem{Gupta2020a}
Harshit Gupta, Michael~T. McCann, Laur{\`{e}}ne Donati, and Michael Unser.
\newblock {CryoGAN: A New Reconstruction Paradigm for Single-particle Cryo-EM
  Via Deep Adversarial Learning}.
\newblock {\em bioRxiv}, page 2020.03.20.001016, mar 2020.

\bibitem{Gupta2020}
Harshit Gupta, Thong~H. Phan, Jaejun Yoo, and Michael Unser.
\newblock {Multi-CryoGAN: Reconstruction of Continuous Conformations in Cryo-EM
  Using Generative Adversarial Networks}.
\newblock {\em Lecture Notes in Computer Science (including subseries Lecture
  Notes in Artificial Intelligence and Lecture Notes in Bioinformatics)}, 12535
  LNCS:429--444, 2020.

\bibitem{harmuth2018mitochondrial}
Tina Harmuth, Caroline Prell-Schicker, Jonasz~J Weber, Frank Gellerich, Claudia
  Funke, Stefan Drie{\ss}en, Janine~CD Magg, Guido Krebiehl, Hartwig Wolburg,
  Stefanie~N Hayer, et~al.
\newblock Mitochondrial morphology, function and homeostasis are impaired by
  expression of an n-terminal calpain cleavage fragment of ataxin-3.
\newblock {\em Frontiers in Molecular Neuroscience}, 11:368, 2018.

\bibitem{7780459}
Kaiming He, Xiangyu Zhang, Shaoqing Ren, and Jian Sun.
\newblock Deep residual learning for image recognition.
\newblock In {\em 2016 IEEE Conference on Computer Vision and Pattern
  Recognition (CVPR)}, pages 770--778, 2016.

\bibitem{Heydari2017}
Tiam Heydari, Maziar Heidari, Omid Mashinchian, Michal Wojcik, Ke Xu,
  Matthew~John Dalby, Morteza Mahmoudi, and Mohammad~Reza Ejtehadi.
\newblock {Development of a Virtual Cell Model to Predict Cell Response to
  Substrate Topography}.
\newblock {\em ACS Nano}, 11(9):9084--9092, sep 2017.

\bibitem{jakobs2020light}
Stefan Jakobs, Till Stephan, Peter Ilgen, and Christian Br{\"u}ser.
\newblock Light microscopy of mitochondria at the nanoscale.
\newblock {\em Annual review of biophysics}, 49:289, 2020.

\bibitem{kanezaki_code}
Asako Kanezaki.
\newblock pytorch-unsupervised-segmentation.
\newblock \url{https://github.com/kanezaki/pytorch-unsupervised-segmentation},
  2018.
\newblock Online; accessed 18-November-2022.

\bibitem{10.1109/ICASSP.2018.8462533}
Asako Kanezaki.
\newblock Unsupervised image segmentation by backpropagation.
\newblock In {\em 2018 IEEE International Conference on Acoustics, Speech and
  Signal Processing (ICASSP)}, page 1543–1547. IEEE Press, 2018.

\bibitem{lefebvre2021automated}
Austin~EYT Lefebvre, Dennis Ma, Kai Kessenbrock, Devon~A Lawson, and Michelle~A
  Digman.
\newblock Automated segmentation and tracking of mitochondria in live-cell
  time-lapse images.
\newblock {\em Nature Methods}, 18(9):1091--1102, 2021.

\bibitem{Li2021}
Ruihui Li, Xianzhi Li, Ka~Hei Hui, and Chi~Wing Fu.
\newblock {SP-GAN: Sphere-guided 3D shape generation and manipulation}.
\newblock {\em ACM Transactions on Graphics}, 40(4), jul 2021.

\bibitem{lichtman2005fluorescence}
Jeff~W Lichtman and Jos{\'e}-Angel Conchello.
\newblock Fluorescence microscopy.
\newblock {\em Nature Methods}, 2(12):910--919, 2005.

\bibitem{lihavainen2012mytoe}
Eero Lihavainen, Jarno M{\"a}kel{\"a}, Johannes~N Spelbrink, and Andre~S
  Ribeiro.
\newblock Mytoe: automatic analysis of mitochondrial dynamics.
\newblock {\em Bioinformatics}, 28(7):1050--1051, 2012.

\bibitem{liu2018learning}
Shikun Liu, Lee Giles, and Alexander Ororbia.
\newblock Learning a hierarchical latent-variable model of 3d shapes.
\newblock In {\em International Conference on 3D Vision (3DV)}, pages 542--551.
  IEEE, 2018.

\bibitem{liu2022multi}
Tianyan Liu, Till Stephan, Peng Chen, Jingting Chen, Dietmar Riedel, Zhongtian
  Yang, Stefan Jakobs, and Zhixing Chen.
\newblock Multi-color live-cell sted nanoscopy of mitochondria with a gentle
  inner membrane stain.
\newblock {\em bioRxiv}, 2022.

\bibitem{Occupancy_Networks}
Lars Mescheder, Michael Oechsle, Michael Niemeyer, Sebastian Nowozin, and
  Andreas Geiger.
\newblock Occupancy networks: Learning 3d reconstruction in function space.
\newblock In {\em IEEE Conference on Computer Vision and Pattern Recognition
  (CVPR)}, 2019.

\bibitem{michalkiewicz1901deep}
M Michalkiewicz, JK Pontes, D Jack, M Baktashmotlagh, and AP Eriksson.
\newblock Deep level sets: Implicit surface representations for 3d shape
  inference. corr abs/1901.06802 (2019), 1901.

\bibitem{Milletari2016VNetFC}
Fausto Milletari, Nassir Navab, and Seyed-Ahmad Ahmadi.
\newblock V-net: Fully convolutional neural networks for volumetric medical
  image segmentation.
\newblock {\em International Conference on 3D Vision (3DV)}, pages 565--571,
  2016.

\bibitem{miyazono2018uncoupled}
Yoshihiro Miyazono, Shingo Hirashima, Naotada Ishihara, Jingo Kusukawa,
  Kei-ichiro Nakamura, and Keisuke Ohta.
\newblock Uncoupled mitochondria quickly shorten along their long axis to form
  indented spheroids, instead of rings, in a fission-independent manner.
\newblock {\em Scientific Reports}, 8(1):1--14, 2018.

\bibitem{ong2010mitochondrial}
Sang-Bing Ong and Derek~J Hausenloy.
\newblock Mitochondrial morphology and cardiovascular disease.
\newblock {\em Cardiovascular Research}, 88(1):16--29, 2010.

\bibitem{pape2020multicolor}
Jasmin~K Pape, Till Stephan, Francisco Balzarotti, Rebecca B{\"u}chner, Felix
  Lange, Dietmar Riedel, Stefan Jakobs, and Stefan~W Hell.
\newblock Multicolor 3d minflux nanoscopy of mitochondrial micos proteins.
\newblock {\em Proceedings of the National Academy of Sciences},
  117(34):20607--20614, 2020.

\bibitem{Park_2019_CVPR}
Jeong~Joon Park, Peter Florence, Julian Straub, Richard Newcombe, and Steven
  Lovegrove.
\newblock Deepsdf: Learning continuous signed distance functions for shape
  representation.
\newblock In {\em IEEE Conference on Computer Vision and Pattern Recognition
  (CVPR)}, June 2019.

\bibitem{Parlakgul2022}
G{\"{u}}neş Parlakg{\"{u}}l, Ana~Paula Arruda, Song Pang, Erika Cagampan, Nina
  Min, Ekin G{\"{u}}ney, Grace~Yankun Lee, Karen Inouye, Harald~F. Hess,
  C.~Shan Xu, and G{\"{o}}khan~S. Hotamışlıgil.
\newblock {Regulation of liver subcellular architecture controls metabolic
  homeostasis}.
\newblock {\em Nature}, 603(7902):736--742, mar 2022.

\bibitem{raju2022deep}
Ashwin Raju, Shun Miao, Dakai Jin, Le Lu, Junzhou Huang, and Adam~P Harrison.
\newblock Deep implicit statistical shape models for 3d medical image
  delineation.
\newblock In {\em AAAI Conference on Artificial Intelligence}, volume~36, pages
  2135--2143, 2022.

\bibitem{ronneberger2015u}
Olaf Ronneberger, Philipp Fischer, and Thomas Brox.
\newblock U-net: Convolutional networks for biomedical image segmentation.
\newblock In {\em International Conference on Medical image computing and
  computer-assisted intervention}, pages 234--241. Springer, 2015.

\bibitem{samanta2019fluorescent}
Soham Samanta, Ying He, Amit Sharma, Jiseon Kim, Wenhui Pan, Zhigang Yang, Jia
  Li, Wei Yan, Liwei Liu, Junle Qu, et~al.
\newblock Fluorescent probes for nanoscopic imaging of mitochondria.
\newblock {\em Chem}, 5(7):1697--1726, 2019.

\bibitem{schindelin2012fiji}
Johannes Schindelin, Ignacio Arganda-Carreras, Erwin Frise, Verena Kaynig, Mark
  Longair, Tobias Pietzsch, Stephan Preibisch, Curtis Rueden, Stephan Saalfeld,
  Benjamin Schmid, et~al.
\newblock Fiji: an open-source platform for biological-image analysis.
\newblock {\em Nature methods}, 9(7):676--682, 2012.

\bibitem{schneider2012nih}
Caroline~A Schneider, Wayne~S Rasband, and Kevin~W Eliceiri.
\newblock Nih image to imagej: 25 years of image analysis.
\newblock {\em Nature methods}, 9(7):671--675, 2012.

\bibitem{sekh2021_codes}
Arif~Ahmed Sekh, Ida~S. Opstad, Gustav Godtliebsen, {\AA}sa~Birna Birgisdottir,
  Balpreet~Singh Ahluwalia, Krishna Agarwal, and Dilip~K. Prasad.
\newblock {Physics-based machine learning for subcellular segmentation in
  living cells}.
\newblock \url{https://doi.org/10.5281/zenodo.5017066}.

\bibitem{Sekh2021}
Arif~Ahmed Sekh, Ida~S. Opstad, Gustav Godtliebsen, {\AA}sa~Birna Birgisdottir,
  Balpreet~Singh Ahluwalia, Krishna Agarwal, and Dilip~K. Prasad.
\newblock {Physics-based machine learning for subcellular segmentation in
  living cells}.
\newblock {\em Nature Machine Intelligence}, 3(12):1071--1080, dec 2021.

\bibitem{shin2018pixels}
Daeyun Shin, Charless~C Fowlkes, and Derek Hoiem.
\newblock Pixels, voxels, and views: A study of shape representations for
  single view 3d object shape prediction.
\newblock In {\em Proceedings of the IEEE conference on computer vision and
  pattern recognition}, pages 3061--3069, 2018.

\bibitem{william_silversmith_2021_5535251}
William Silversmith.
\newblock {seung-lab/connected-components-3d: Zenodo Release v1}, Sept. 2021.

\bibitem{SLEPCHENKO2003570}
Boris~M. Slepchenko, James~C. Schaff, Ian Macara, and Leslie~M. Loew.
\newblock Quantitative cell biology with the virtual cell.
\newblock {\em Trends in Cell Biology}, 13(11):570--576, 2003.

\bibitem{stephan2019live}
Till Stephan, Axel Roesch, Dietmar Riedel, and Stefan Jakobs.
\newblock Live-cell sted nanoscopy of mitochondrial cristae.
\newblock {\em Scientific reports}, 9(1):1--6, 2019.

\bibitem{pmlr-v97-tan19a}
Mingxing Tan and Quoc Le.
\newblock {E}fficient{N}et: Rethinking model scaling for convolutional neural
  networks.
\newblock In Kamalika Chaudhuri and Ruslan Salakhutdinov, editors, {\em
  Proceedings of the 36th International Conference on Machine Learning},
  volume~97 of {\em Proceedings of Machine Learning Research}, pages
  6105--6114. PMLR, 09--15 Jun 2019.

\bibitem{thornburg2022fundamental}
Zane~R Thornburg, David~M Bianchi, Troy~A Brier, Benjamin~R Gilbert, Tyler~M
  Earnest, Marcelo~CR Melo, Nataliya Safronova, James~P S{\'a}enz, Andr{\'a}s~T
  Cook, Kim~S Wise, et~al.
\newblock Fundamental behaviors emerge from simulations of a living minimal
  cell.
\newblock {\em Cell}, 185(2):345--360, 2022.

\bibitem{viana2015quantifying}
Matheus~Palhares Viana, Swee Lim, and Susanne~M Rafelski.
\newblock Quantifying mitochondrial content in living cells.
\newblock In {\em Methods in cell biology}, volume 125, pages 77--93. Elsevier,
  2015.

\bibitem{mitograph_code}
Matheus~Palhares Viana, Swee Lim, and Susanne~M Rafelski.
\newblock Mitograph v3.0.
\newblock \url{https://github.com/vianamp/MitoGraph/releases/tag/v3.0}, 2018.
\newblock Online; accessed 18-November-2022.

\bibitem{mitometer_code}
Matheus~Palhares Viana, Swee Lim, and Susanne~M Rafelski.
\newblock Mitometer.
\newblock \url{https://github.com/aelefebv/Mitometer}, 2021.
\newblock Online; accessed 18-November-2022.

\bibitem{wei2020mitoem}
Donglai Wei, Zudi Lin, Daniel Franco-Barranco, Nils Wendt, Xingyu Liu, Wenjie
  Yin, Xin Huang, Aarush Gupta, Won-Dong Jang, Xueying Wang, et~al.
\newblock Mitoem dataset: Large-scale 3d mitochondria instance segmentation
  from em images.
\newblock In {\em International Conference on Medical Image Computing and
  Computer-Assisted Intervention}, pages 66--76. Springer, 2020.

\bibitem{3dgan}
Jiajun Wu, Chengkai Zhang, Tianfan Xue, William~T Freeman, and Joshua~B
  Tenenbaum.
\newblock Learning a probabilistic latent space of object shapes via 3d
  generative-adversarial modeling.
\newblock In {\em Advances in Neural Information Processing Systems}, pages
  82--90, 2016.

\bibitem{Wu_2020_CVPR}
Rundi Wu, Yixin Zhuang, Kai Xu, Hao Zhang, and Baoquan Chen.
\newblock Pq-net: A generative part seq2seq network for 3d shapes.
\newblock In {\em IEEE/CVF Conference on Computer Vision and Pattern
  Recognition (CVPR)}, June 2020.

\bibitem{xiao2018automatic}
Chi Xiao, Xi Chen, Weifu Li, Linlin Li, Lu Wang, Qiwei Xie, and Hua Han.
\newblock Automatic mitochondria segmentation for em data using a 3d supervised
  convolutional network.
\newblock {\em Frontiers in Neuroanatomy}, 12:92, 2018.

\bibitem{NIPS2019_8340}
Qiangeng Xu, Weiyue Wang, Duygu Ceylan, Radomir Mech, and Ulrich Neumann.
\newblock Disn: Deep implicit surface network for high-quality single-view 3d
  reconstruction.
\newblock In H. Wallach, H. Larochelle, A. Beygelzimer, F. d\textquotesingle
  Alch\'{e}-Buc, E. Fox, and R. Garnett, editors, {\em Advances in Neural
  Information Processing Systems 32}, pages 492--502. Curran Associates, Inc.,
  2019.

\bibitem{Yang_2022_CVPR}
Jiancheng Yang, Udaranga Wickramasinghe, Bingbing Ni, and Pascal Fua.
\newblock Implicitatlas: Learning deformable shape templates in medical
  imaging.
\newblock In {\em IEEE Conference on Computer Vision and Pattern Recognition
  (CVPR)}, pages 15861--15871, June 2022.

\bibitem{zamponi2018mitochondrial}
Nahuel Zamponi, Emiliano Zamponi, Sergio~A Cannas, Orlando~V Billoni, Pablo~R
  Helguera, and Dante~R Chialvo.
\newblock Mitochondrial network complexity emerges from fission/fusion
  dynamics.
\newblock {\em Scientific reports}, 8(1):1--10, 2018.

\bibitem{Zhou2020UNetRS}
Zongwei Zhou, Md~Mahfuzur~Rahman Siddiquee, Nima Tajbakhsh, and Jianming Liang.
\newblock Unet++: Redesigning skip connections to exploit multiscale features
  in image segmentation.
\newblock {\em IEEE Transactions on Medical Imaging}, 39:1856--1867, 2020.

\end{thebibliography}
}
\pagebreak

\twocolumn[{%
\renewcommand\twocolumn[1][]{#1}%
\begin{center}
\large Supplemental Materials for \\ \textbf{ MiShape: 3D Shape Modelling of Mitochondria in Microscopy}
\end{center}
}]

\renewcommand{\thesection}{S\arabic{section}} 
\setcounter{section}{0}
\renewcommand{\thetable}{S\arabic{table}}  
\setcounter{table}{0}
\renewcommand{\thefigure}{S\arabic{figure}}
\setcounter{figure}{0}
\renewcommand{\thefigure}{S\arabic{figure}}
\setcounter{figure}{0}

\section{Microscope-to microscope transformation} \label{sec:M-to-M}
Here, we consider the problem of using 2D fluorescence images or z-stack generated by one fluorescence microscope to simulate the microscopy images or z-stack corresponding to another fluorescence microscope. 

\begin{figure*}
    \centering
    \includegraphics[width=1\linewidth]{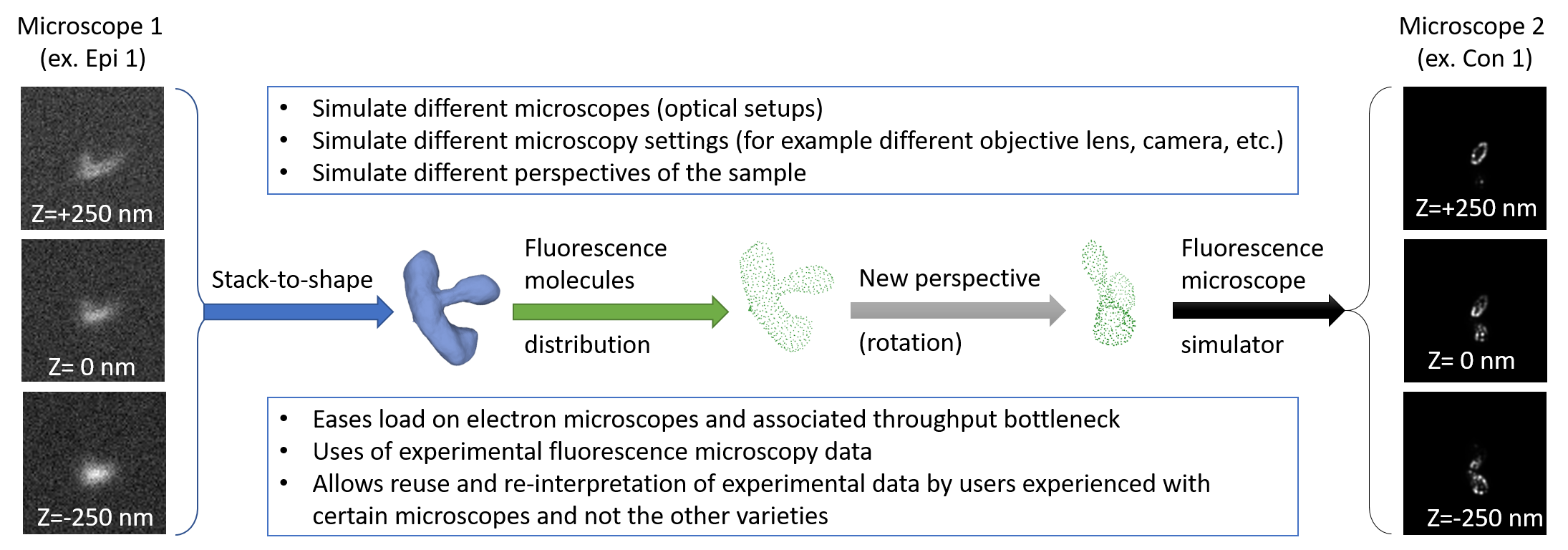}
    \caption{The concept of microscope-to-microscope transformation illustrated using an example. }
    \label{fig:M-to-M}
\end{figure*}

Having generated the 3D implicit shapes using the 3D EM images, we presented the methodology to convert the implicit shapes to fluorescence microscopy images in Sec.  
\ref{sub_sec:EM2Fl} 
of the main paper. 
Further, in Sec.  
\ref{sub_sec:3D from 2D} 
we presented the methodology to derive 3D shapes from 2D fluorescence images or fluorescence z-stack. For the purpose of fluorescence microscope-to-microscope translation, we propose to combine these works as follows. First, we use the fluorescence image or z-stack from the source microscope to construct the corresponding 3D shape. Then, we scale the 3D shape to the actual physical units. The scale factor can be computed through the minimization approach described in the last paragraph of 
\ref{sub_sec:3D from 2D}
if it is unknown. Here, we assume that the scale factor is known since we intend to illustrate the concept and feasibility. 
Having brought the implicit shape into the relevant physical dimensions, it may be used as it is or subject to rotation, translation, or creation of collage of many such structures. Then the process of generating the fluorescence molecule distribution and the fluorescence images or z-stacks presented in Sec.  
\ref{sub_sec:EM2Fl} 
are applied to generate the images. 
Fig. \ref{fig:M-to-M} outline the steps in the top panel and show the results in the bottom panel. 

Such a transformation is useful to avoid using simulation-only datasets for AI, and incorporating experimental evidence without resorting to low throughput and expertise-demanding EM imaging. Moreover, experiments in biology are conducted over timelines of months to years. In this span, it often happens that imaging has to be done on different microscopes considering the simultaneous availability of microscopes and biological samples. This often prohibits one-to-one comparison of biological images related to the same experiment. This problem can be solved through the proposed approach. Similarly, it can help in situations where the imaging was done in one microscopy setting (for example one objective lens) but it is desirable to study the image at another setting. 

Further, it can be used to support the reinterpretation of the data by a biologist more used to one kind of microscope which is different from the microscope from which the data was acquired. This is useful because the biologists learning to interpret data from one microscope also learn the limitations of the microscope and how these should be incorporated in interpretation and deriving inferences about biological questions. However, this abstract heuristic knowledge is not easily transferable across microscopes. Also, it can reduce the dependence on relatively lower throughput, costly, and heavily used confocal microscopes by forming confocal image equivalents using cheaper and high throughput epifluorescence microscopes. 


\begin{figure*}
    \centering
    \includegraphics[width=1\linewidth]{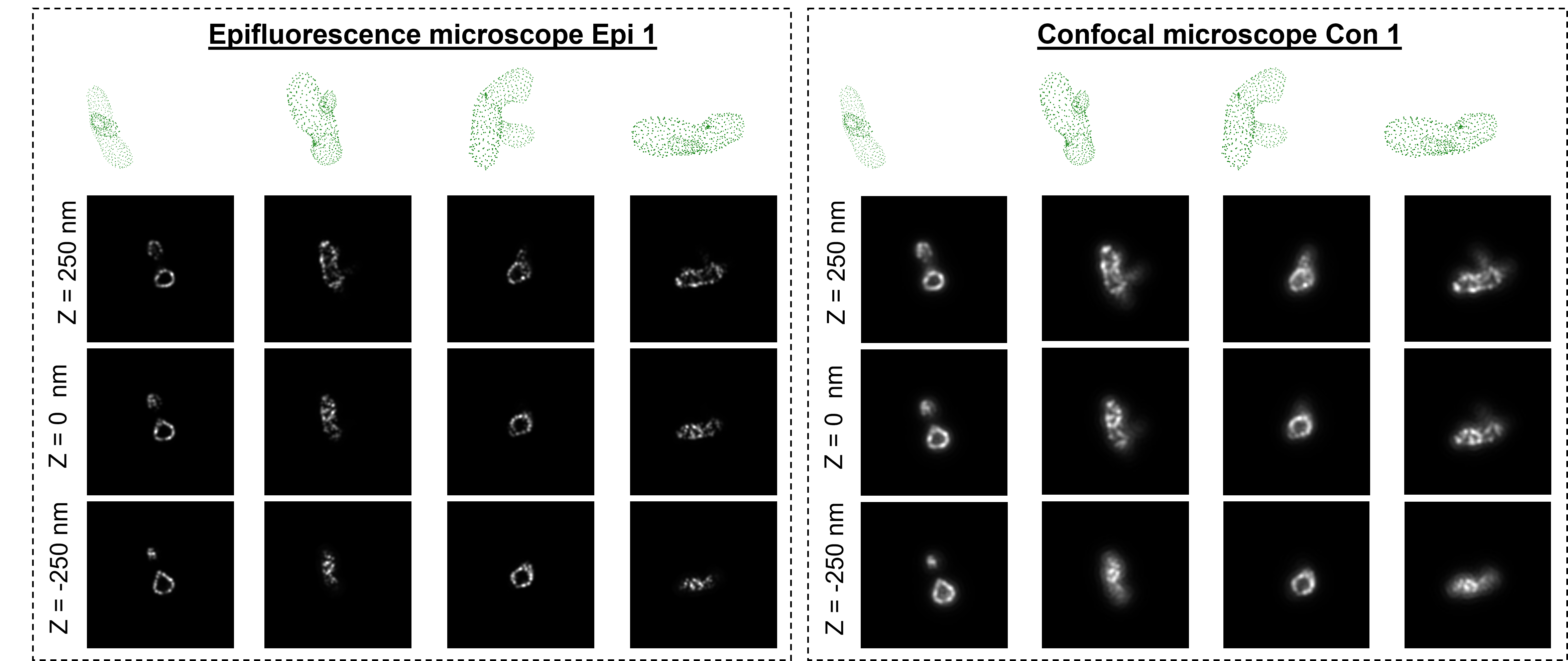}
    \caption{Microscope-to-microscope transformation results for the shape used in Fig. \ref{fig:M-to-M}. Simulated (noise-free) images created for two different microscopes using different perspectives of the shapes learnt by the stack-to-shape model are presented.}
    \label{fig:M-to-M-examples}
\end{figure*}

 
\section{Microscope Parameters - simulated and experimental}
We have simulated two epifluorescence microscopes and one confocal microscope in this paper. Table \ref{tab:microscopes} lists the optical parameters of these microscopes. It also includes the sections in the main manuscript in which these microscopes were used. We note that Epi2 was simulated to match the experimental microscopy data used in 2D segmentation \cite{sekh2021_codes}.
 
The publicly available source provided in \cite{sekh2021_codes} was used for simulating the fluorescence microscopy images. Noise was simulated on top of all the simulated microscopy images to emulate a signal-to-background ratio between  2 to 4. For this, the value of background $b$ was set around 100. 
 
 \begin{table*}
     \centering
     \begin{tabular}{l|c c c}
     \hline
                                &  \textbf{Con1 }         & \textbf{Epi1}            & \textbf{Epi2} \\
        \hline \hline
        \multicolumn{4}{l}{\textbf{{Experiments}}}\\
                                            \hline
       EM to fluorescence image (sec. \ref{sub_sec:res_Ep-to-Fl}, Fig. \ref{fig:Fig6_em}) &   Fig. \ref{fig:res_Ep-to-F} &  Fig. \ref{fig:res_Ep-to-F} & $\times$\\
       Stack to shape (sec. \ref{sub_sec:res_2D_to_3D}) & $\times$ &  Figs. \ref{fig:CompareFl2EM}, \ref{fig:Epistack_to_3d}, Table \ref{tab:shape_from_fluorescence}  & $\times$\\
       Image to shape (sec \ref{sub_sec:res_2D_to_3D}) &  Figs. \ref{fig:CompareFl2EM}, \ref{fig:Confocal2d_to_3d}, \ref{fig:Epi_image-to-shape}, Table \ref{tab:shape_from_fluorescence} &  Figs. \ref{fig:CompareFl2EM}, \ref{fig:Epi_image-to-shape}, Table \ref{tab:shape_from_fluorescence} & $\times$\\
       2D segmentation (sec \ref{sub_sec:res_2Dsegmenation}) & $\times$ & $\times$ &  Fig. \ref{fig:seg_res}, \ref{fig:2d_seg_samples}, Tables \ref{table:our_vs_arif}, \ref{table:benchmark} \\
       Microscope-to-microscope (sec. \ref{sec:M-to-M}) & Figs. \ref{fig:M-to-M},\ref{fig:M-to-M-examples} & Fig. \ref{fig:M-to-M},\ref{fig:M-to-M-examples} & $\times$\\
\hline \hline
\multicolumn{4}{l}{\textbf{Microscope details}}\\ \hline
        Type of microscope      & Confocal      & Epifluorescence  & Epifluorescence \\
        Emission wavelength     &  600 nm       &  688 nm         & 608nm \\
        Pixel Size              &  70 nm        & 109 nm           & 80 nm  \\
        Numerical Aperture      &  1.4          & 1.42             & 1.4  \\
        Magnification           &  63           &  60              & 60  \\
        Optical Resolution (nm) &  152             & 245                 & 217  \\
        Optical Resolution (pixels)  & 2.16          & 2.22                 & 2.71   \\ \hline
     \end{tabular}
     \caption{Microscope Settings used for various experiments. }
     \label{tab:microscopes}
 \end{table*}

 \paragraph{Deriving emitter locations from implicit shapes}
We noted in the dataset description in section 4 that we downsampled the EM images from the original 8 nm resolution to the more computationally amenable 24 $nm$ resolution. In Sec. \ref{sub_sec:EM2Fl} and Sec. \ref{sub_sec:res_Ep-to-Fl} pertaining to EM to fluorescence microscopy image generation, we sample points on the mesh surface as possible emitter locations. In experiment 4.2, implicit shapes are not used. However, in the future, it will be beneficial to use implicit shapes derived from the EM images preferably at the lowest resolution possible. Having formed the implicit shapes, we could derive meshes at any desired resolution, dependent on, for example, the emitter density or the distribution of host protein on which the fluorescent molecules are expected to bind, or another experimental condition that influences the fluorescence molecule distribution. The implicit shapes also allow the creation of surface or volume meshes. This would therefore allow a much richer and more flexible paradigm for the emulation of conditions encountered in biological experiments.

\section{Architecture and implementation details - Unconditional Generation, Image-to-shape, and Stack-to-shape Reconstruction}
We outline the architectures of the unconditional shape generative model and image-to-shape models in Fig. \ref{fig:sup_arch} and Fig. \ref{fig:sup_arch_cond} respectively. We use the architecture of the original Occupancy Network \cite{Occupancy_Networks} for the unconditional, and the Image-to-Shape models. The occupancy encoder and decoder for all three models explained below have the same architecture. 

\paragraph{Dataset and Splits }
The dataset used for the experiments is the individual components extracted from the segmentation masks available from \cite{Parlakgul2022}. We take a volume of shape 9700 x 9650 x 3629 having an isotropic resolution of 8 $nm$ and re-sample this to a volume of 24 $nm$. 27272 individual shape instances extracted from this are converted to implicit representation following the steps described in \cite{Occupancy_Networks} and the dataset is split as 70-10-20 percent of the whole between the training, validation, and test subsets.

\subsection{Unconditional generative model}
\begin{figure*}
    \centering
    \includegraphics[width=0.85\linewidth]{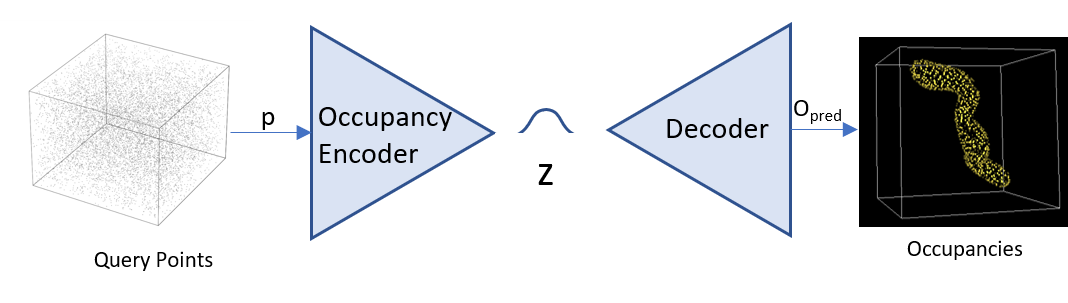}
    \caption{Unconditional generative model used to learn the implicit shapes of Mitochondria. The details of the Occupancy Encoder and Decoder is discussed in section  }
    \label{fig:sup_arch}
\end{figure*}

The architecture of this model is outlined in Fig. \ref{fig:sup_arch}.
The occupancy encoder receives the batches of query points $p$ as input. The encoder-decoder modules form a VAE-like construction. The occupancy encoder consists of a stack of four fully connected layers, followed by two fully-connected layers to produce both the mean and log of the standard deviation of the 128-dimensional latent code $z$. The decoder consists of five fully connected ResNet blocks. The final layer of the decoder is a fully connected layer of output size one. The decoder predicts the occupancies of the query points and the model is trained on a binary cross-entropy loss between the predicted occupancies and true occupancies. 

\subsection{Image-to-shape Reconstruction model}
\begin{figure*}
    \centering
    \includegraphics[width=.85\linewidth]{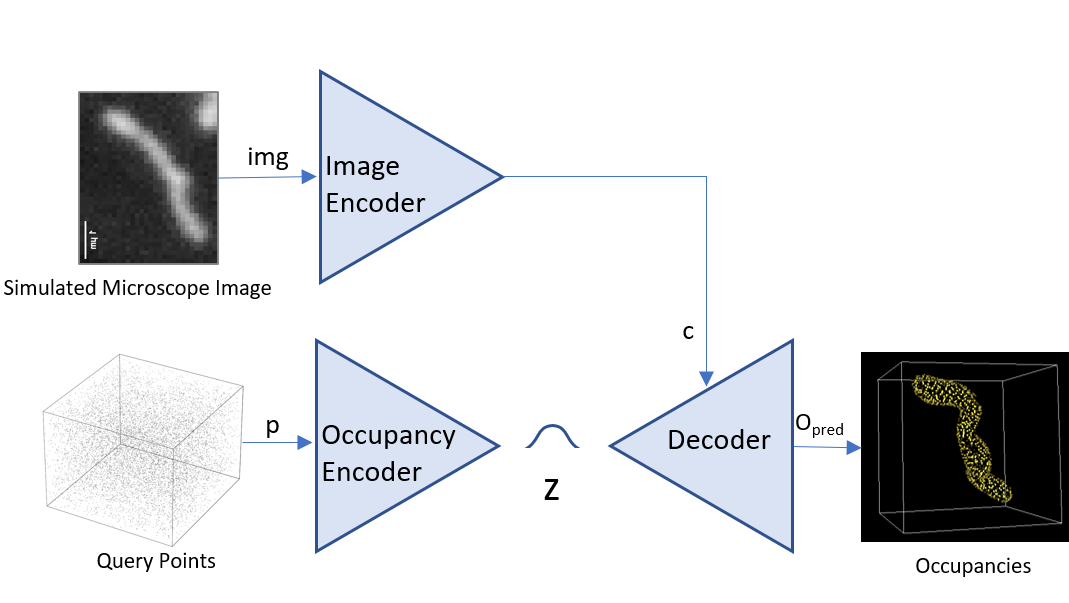}
    \caption{Conditional generative model used to learn the implicit shapes of Mitochondria}
    \label{fig:sup_arch_cond}
\end{figure*}

The architecture of image-to-shape reconstruction models is outlined in Fig. \ref{fig:sup_arch_cond}.
In this mode, we use the simulated microscope image of the input shape to condition the decoder. The simulated image is input to a 2D ResNet- 18 \cite{7780459} to produce the conditioning $c$. The conditioning with the image input is done using conditional batch normalization on each of the ResNet blocks of the decoder. 

\paragraph{Metrics}
The metric used for the evaluation of the image-to-shape and the stack-to-shape models are volumetric IoU and Chamfer-$L_1$ distance. If $M_{pred}$ and  $ M_{gt}$ are the set of all points that are inside or on the mesh of the predicted and the ground truth mesh, \cite{Occupancy_Networks} defines volumetric IoU as, 

\begin{equation}
    IoU( M_{pred}, M_{gt} ) \equiv \frac{  | {M_{pred} \cap M_{gt}  } |}{ { | M_{pred} \cup M_{gt} | } }
    \label{eq:viou}
\end{equation}
 and the Chamfer-$L_1$ distance is defined as,

\begin{align}
\begin{split}
    \text{Chamfer}{\text -} & L_1( M_{pred}, M_{gt} ) \equiv \\
  &  \frac{ 1}{  2 | \partial M_{pred} | } \int_{\partial M_{pred}} \min_{q\in M_{gt}} || p-q|| dp  +  \\
 &\frac{ 1}{  2 | \partial M_{gt} | } \int_{\partial M_{gt}} \min_{q\in M_{pred}} || p-q|| dq
\end{split}
    \label{eq:chamfer}
\end{align}

\subsection{Stack-to-shape Reconstruction model}
 The stack-to-image model follows the same basic architecture as the image-to-shape reconstruction model but the encoder is a 3D-ResNet \cite{Feichtenhofer_2019_ICCV} module in place of a 2D-ResNet used for image-to-shape reconstruction.

\section{EM data used in Fig. 6 for generating fluorescence microscopy images}

\begin{figure*}
    \centering
    \includegraphics[width=1\linewidth]{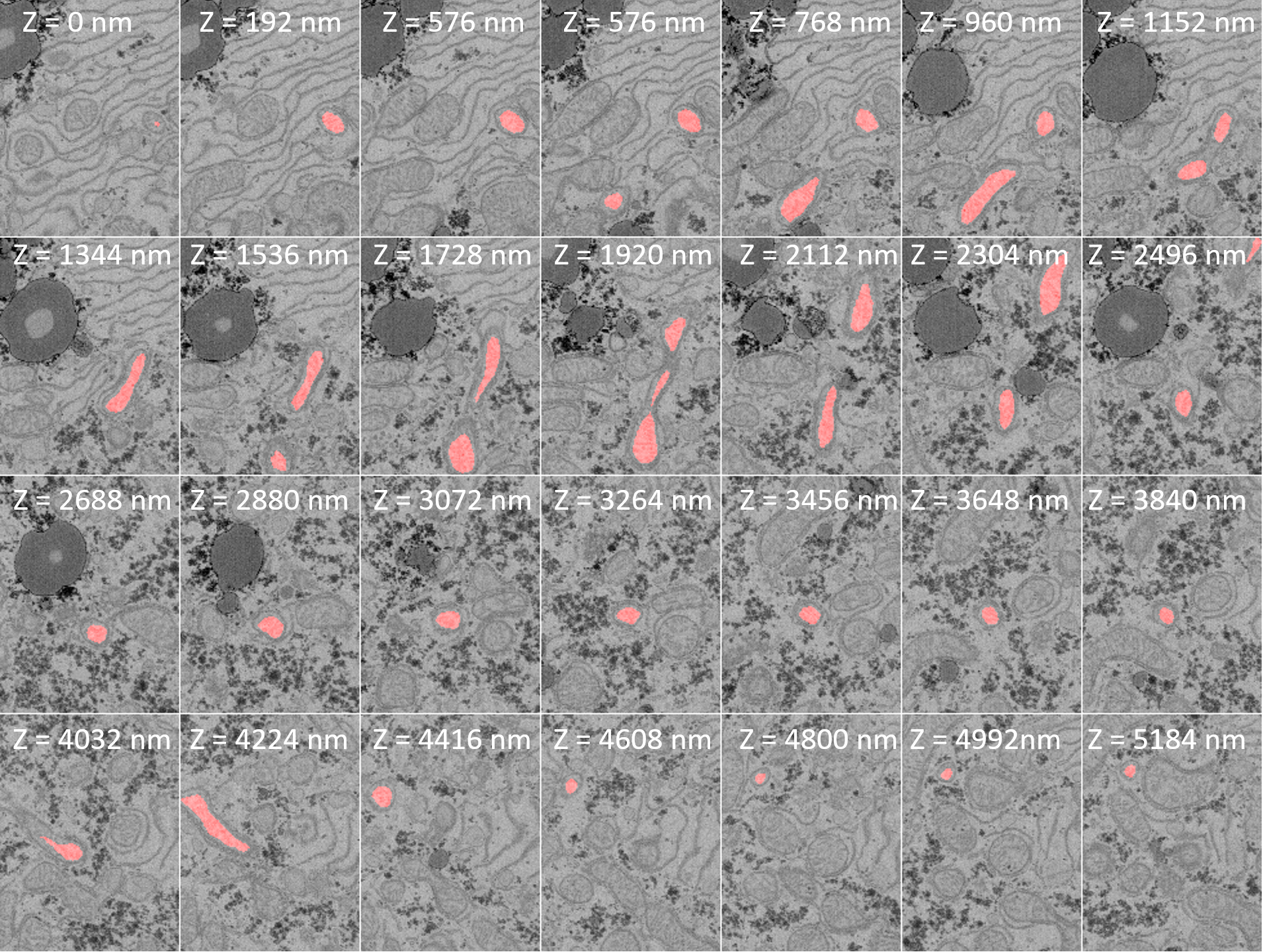}
    \caption{EM images used to generate the fluorescence microscopy images shown in Fig. 6 of the main manuscript.}
    \label{fig:Fig6_em}
\end{figure*}

The EM data used for generating the fluorescence microscopy images is shown in Fig. \ref{fig:Fig6_em}. The montage shows slices sampled at $192nm$.  


\section{Qualitative results for epifluorescence image-to-stack experiment}
\begin{figure}
    \centering
    \includegraphics[width=1\linewidth]{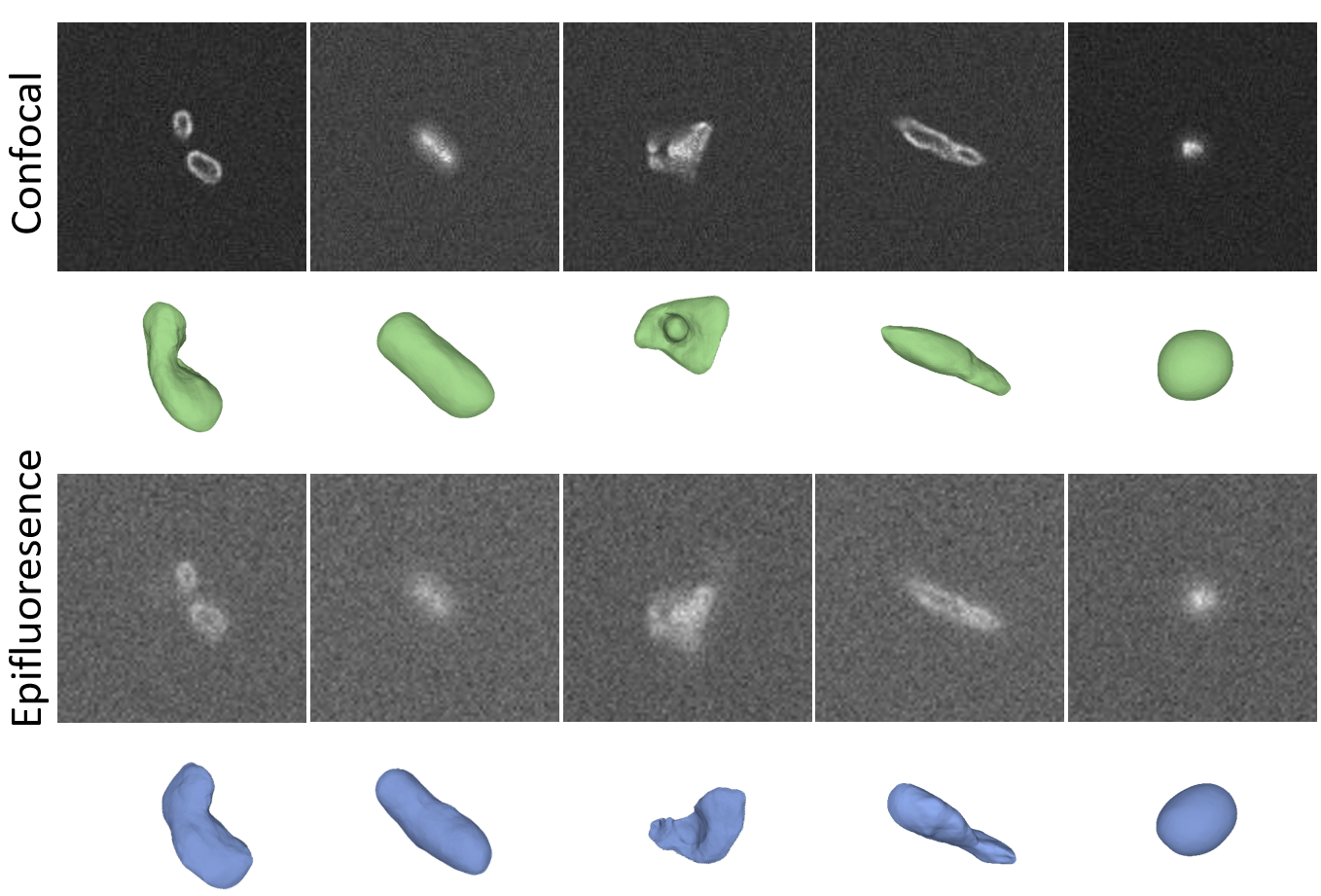}
    \caption{Qualitative results for epifluorescence image-to-shape and comparison with confocal image-to-shape and epifluorescence stack-to-shape}
    \label{fig:Epi_image-to-shape}
\end{figure}

We provide qualitative results of the epifluorescence image-to-stack experiment in Fig. \ref{fig:Epi_image-to-shape} with a comparison to the confocal-image-to-stack counterparts of the experiment. The images simulated for this experiment are using the Epi1 configuration. 


\section{2D segmentation: Experimental Details}
\begin{figure}
    \centering
    \includegraphics[width=1\linewidth]{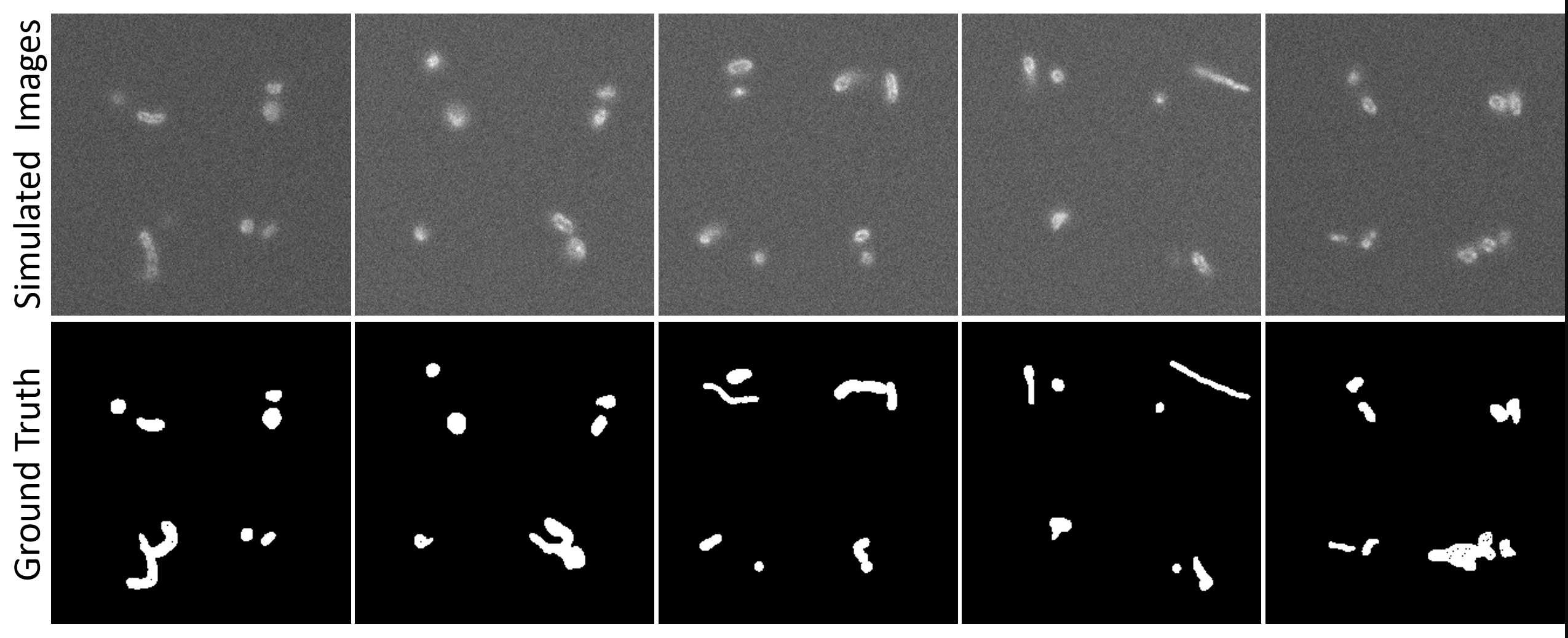}
    \caption{Examples of our simulated training dataset created using EM data for 2D Segmentation. }
    \label{fig:2d_seg_samples}
\end{figure}

\paragraph{Simulation-supervised training dataset}\label{seb_sec:our_sim_seg}
The individual mitochondrial shapes extracted are used to create a training dataset for the task of 2D segmentation.
We refer to the source \cite{sekh2021_codes} for creating the simulated training dataset. Images are created with 2 mitochondria in a $128 \times 128 $ image. Four such images are montaged together to form a $256 \times 256$ sized image that is the size of the input for the segmentation model used in \cite{Sekh2021}. We are currently unable to simulate larger mitochondrial shapes as the extraction of ground truth is challenging for technical reasons related to computational resources. Examples of the dataset is shown in Fig. \ref{fig:2d_seg_samples}. We create a dataset of 7000 image-ground truth pairs and evaluate the feasibility of using the extracted shapes for creating realistic datasets. 

%
\paragraph{Architecture} The segmentation model from \cite{Sekh2021} uses an EfficientNet\cite{pmlr-v97-tan19a} backbone and is trained for 20 epochs with early stopping. The learning rate is set to 1e-4 and the batch size to 32. We train two models: one that is trained using only the simulated dataset and another that is first pre-trained on the simulated dataset and then fine-tuned to the training split of the manually annotated dataset from \cite{Sekh2021}.  The models fine-tuned on the real microscope hand-labeled training images are indicated using TL (Transfer Learning). We benchmark the performance of these models and other popular methods used for fluorescence mitochondria image segmentation in Table \ref{table:benchmark}.


\subsection{2D segmentation baselines}
We discuss the working and configuration of each of the baseline methods used to benchmark the task of 2D segmentation of mitochondria in Epifluorescence images, in Table \ref{table:benchmark}. 

\paragraph{MitoGraph} \cite{viana2015quantifying} is a fully automated image-processing software, designed for mitochondrial analysis. It is mainly used for calculating their 3D volume and creating a graph representation for ease of analysis. It takes as input the max-intensity projection of an image stack and produces a segmentation map in addition to a 3D surface and a graph structure. For the segmentation part, the authors use a modified multiscale vesselness filter \cite{frangi1998multiscale} to separate mitochondria from the background. The output of this filter is a matrix, where the value of each voxel denotes the probability of it belonging to a mitochondria. The normalized gradient of this matrix is then converted to the final segmentation output using a threshold.

We used MitoGraph software V3.0 \cite{mitograph_code} which added support for 2D input data. For the test dataset, all the optional arguments were set to their default values, and the pixel size was set to 0.08 microns to match our test dataset's. 

\paragraph{Mitometer} \cite{lefebvre2021automated} is a fully automated segmentation and tracking algorithm designed for mitochondrial analysis. It works on both 2D or 3D images provided as a time-lapse stack. The authors utilize a diffused background removal step which involves applying a set of circular median filters of varying sizes and finding the lowest value for each pixel. This produces a 'diffuse' background, a rough estimation of the actual background, which is then removed from the input image. Doing this reduces the background noise while preserving the shape and size of the mitochondria in foreground. The background-removed input image is then passed through a gaussian kernel and intensity-thresholded using Otsu's threshold to obtain the final segmentation output.

The Mitometer software \cite{mitometer_code} requires pixel size and time between frames as input. For our single-frame 2D images, we set pixel size to 80 $nm$ (pixel size of the dataset) and kept the time between frames to its default value of 1 second. While Mitometer also produces an instance segmentation mask, we considered only the final output, which combines all the instance masks into one single segmentation, which we need for this comparison.
As a side note, we observed that the software had difficulty processing some of the test images with very small segmentation masks and would crash on including these images. The problematic images were thus removed from the test dataset and were not used for any of the methods studied here.

%
%

\paragraph{MitoSegNet} \cite{fischer2020mitosegnet} utilizes a modified U-Net \cite{ronneberger2015u} architecture for segmenting mitochondria. The modifications help the model train faster than the vanilla U-Net, according to the authors. MitoSegNet was trained on a small in-house dataset of mitochondria in body wall muscle cells of adult C. elegans worms.

We used their publicly released software \cite{mitosegnet_code} with the pre-trained model \cite{mitosegnet_model} provided by the authors. We used the 2D option in the GUI with the minimum object size (in pixels) set to its default value of zero. Setting this to higher values did not result in an increase in the score.

The performance of MitoSegNet could potentially be improved using transfer learning. However, given the limited time and the significantly poorer performance of baseline MitoSegNet in comparison to our baseline model and PhySeg (Table \ref{table:benchmark}), we have not performed transfer learning on MitoSegNet.

\paragraph{PhySeg} \cite{Sekh2021} We refer to the simulation supervised segmentation of mitochondria from \cite{sekh2021_codes} as PhySeg. PhySeg simulates training data of fluorescence mitochondria images and ground truth. The shapes of the mitochondria are obtained considering a volume with a fixed radius around parametric curves in 3D. The geometry generation is followed by emitter sampling on the surface, PSF convolution and noise addition. The convolved images are also binarized to obtain the ground truth masks. We refer to the segmentation model and steps of creating the training dataset from this work and hence, it is the same as explained above in section \ref{seb_sec:our_sim_seg}. 

\paragraph{Weka Trainable Segmentation} \cite{10.1093/bioinformatics/btx180} is a segmentation toolbox focused on ease of use. It contains a number of classifiers that can be trained on a single image and then applied to a set of images. For training the classifier, it requires input from user in form of scribbles of different that denote pixels belonging to different classes. The classifier is trained on these pixels and can be then used for inference on other images.

Weka can be installed as a plugin to the image processing software ImageJ \cite{schneider2012nih}. For the purpose of this analysis, we used Fiji \cite{schindelin2012fiji}, which is based on ImageJ and comes pre-installed with Weka. We used the default classifier present in Weka, which is a random forest classifier. All the classifier specific arguments were set to their default values. The classifier was trained by highlighting the mitochondria with class 1 and highlighting the background as class 2, with small scribble interactions. This classifier was then used to generate predictions for all of the test images.

%
%

\paragraph{Kanezaki et. al. 2018} \cite{10.1109/ICASSP.2018.8462533} utilises self-training to achieve unsupervised segmentation using deep learning. The input image pixels are assigned fixed, random labels at first. As the training progresses, the feature extraction network produces per pixel features which are compared to refine the pixel-wise labels. Pixels with similar features get assigned similar labels after each iteration. These newly assigned labels are now used as the ground truth for the next iteration of training. After a couple iterations, the model eventually converges to a stable segmentation map.

We used the publicly available code \cite{kanezaki_code} from the authors and applied the provided demo script with number of iterations set to the default value of 1000, and the minimum number of labels set to 2 (background and mitochondria). Since the labels produced by this method are not the same every time, we manually went through each of the predictions and chose the label value that had the best overlap with the ground truth as the final predictions.


\subsection{Test dataset and Quantitative Results}
We tested our method against a set of established segmentation methods that are tailored for mitochondria. The methods were tested on a test dataset consisting of a total of 107 hand-labelled microscopy images of mitochondria. This test dataset is a subset of the test dataset used in \cite{sekh2021_codes}. Dice and IoU (intersection over union) scores were calculated for each pair of predicted and ground truth images. Dice score can be unstable when comparing a large number of zero pixels. To counter this, only the foreground pixels were compared in the calculation of dice score to reduce the influence of zero pixels in the background.

Table \ref{table:benchmark} tabulates the obtained results. We observe that the simulation-supervised methods perform better than all other methods including image processing-based, interaction-supervised, unsupervised, and even deep learning based methods, in terms of both dice score and IoU. Even though MitoSegNet is a deep learning-based segmentation model, it is not extendable for fine-tuning with another dataset. PhySeg and our model do not differ in architecture, but only on the data that is trained on. We see that PhySeg performs better than ours (row 6 vs row 7 ) without fine-tuning on the hand-labeled training set, and the opposite trend is observed (row 7 vs row 8) with the fine-tuning. We attribute the better generalization of our model to the presence of complex shapes of mitochondria in our simulated training dataset. We investigate this further, with further experiments on simulated test sets from both the works in the following Sec. \ref{sub_sec:physeg_comp}

\begin{table*}
\centering
\begin{tabular}{p{4cm} p{5cm} p{3cm} p{3cm}}
\textbf{Method }                         & \textbf{Type}                 & \textbf{Dice Score $\pm$ std}        & \textbf{IoU $\pm$ std} \\ \hline
MitoGraph \cite{viana2015quantifying}   & Image Processing & 0.586 $\pm$ 0.166         & 0.430 $\pm$ 0.138       \\
Mitometer \cite{lefebvre2021automated}  & Image Processing & 0.665 $\pm$  0.149         & 0.513  $\pm$  0.136        \\
Weka \cite{10.1093/bioinformatics/btx180}        & Interaction Supervised & 0.471 $\pm$  0.179        & 0.325  $\pm$  0.145       \\
Kanezaki et. al. of 2018 \cite{10.1109/ICASSP.2018.8462533}  &   Unsupervised Learning       & 0.560 $\pm$  0.208      & 0.414 $\pm$  0.178      \\
MitoSegNet \cite{fischer2020mitosegnet} & Deep learning       & 0.630 $\pm$ 0.165    & 0.478 $\pm $ 0.149 \\
PhySeg \cite{Sekh2021}                  & Simulation Supervised  & 0.889 $\pm$  0.055   & 0.804  $\pm$ 0.082 \\

Ours                                    & Simulation Supervised  & 0.766 $\pm$  0.131         & 0.635  $\pm$  0.142        \\
PhySeg $+$ TL \cite{Sekh2021}          & Simulation Supervised   & 0.859 $\pm$  0.057         & 0.757  $\pm$  0.080          \\
Ours $+$ TL                             & Simulation Supervised    & 0.864 $\pm$  0.075        & 0.767  $\pm$  0.088       \\

\end{tabular}
\caption{\label{table:benchmark} A quantitative comparison of the segmentation methods tested on the test dataset. Values are averaged over all the test images and the number in the parentheses denotes the standard deviation. (TL indicates transfer learning done on the hand-labeled training dataset from the real microscope images.  )}
\end{table*}

\subsection{Comparison with PhySeg} \label{sub_sec:physeg_comp}

\begin{table*}
\centering
\begin{tabular}{ p{1.5 cm} p{2cm} p{3cm} p{2.5cm} p{2.5cm} p{2.5cm} }
\hline
&
\textbf{Model} & \textbf{Train Data} & \textbf{Test Data}      & \textbf{Dice Score $\pm$ std}  & \textbf{IoU $\pm$ std}   \\ \hline\hline
\textbf{Case 1} & Ours   &  Ours & Ours & 0.860 $\pm$ 0.042 & 0.756 $\pm$ 0.061 \\
& PhySeg & PhySeg & Ours & 0.810 $\pm$ 0.038 & 0.683 $\pm$ 0.053\\
\hline
\textbf{Case 2} & Our & Ours & PhySeg  & 0.835 $\pm$ 0.040 & 0.719 $\pm$ 0.058 \\
& PhySeg & PhySeg & PhySeg & 0.939 $\pm$ 0.014 & 0.885 $\pm$ 0.024 \\
\hline
\textbf{Case 3} & Ours   & Our & Manual & 0.766 $\pm$ 0.131 & 0.635 $\pm$ 0.142 \\
& PhySeg & PhySeg & Manual & 0.889 $\pm$ 0.055 & 0.804 $\pm$ 0.082\\
\hline
\textbf{Case 4} & Ours + TL  & Ours + Manual &  Manual & 0.864 $\pm$ 0.075 & 0.767 $\pm$ 0.088 \\
& PhySeg + TL & PhySeg + Manual & Manual & 0.859 +- 0.057 & 0.757 +- 0.080\\
\hline
\end{tabular}
\caption{\label{table:our_vs_arif} Performance comparison of 2D segmentation with training data from our work and PhySeg. (TL indicates transfer learning done on the hand-labeled training dataset from the real microscope images.  ) }
\end{table*}

We performed comparative experiments between our model and PhySeg \cite{Sekh2021} using different test datasets in order to understand the advantages and disadvantages of both the methods. 
These results are presented in Table \ref{table:our_vs_arif}. 

We have considered 4 cases, where each case uses a specific test dataset. Case 1 uses our simulation supervised dataset created using the shapes learnt through the EM data. Case 2 uses simulation supervised dataset created using the tubular geometry simulation reported in PhySeg. Case 3 uses the manually annotated experimental test dataset of PhySeg. Case 4 is similar to case 3, but in case 4 the models are fine tuned on the manually annotated experimental training dataset shared of PhySeg.

In each case, the PhySeg model was trained on simulation supervised dataset created using the tubular geometry simulation reported in PhySeg and our model was trained on our simulation supervised dataset created using the shapes learnt through the EM data. 

The results of case 1 show that our model performs better than PhySeg. This is because PhySeg was trained on dataset with simple tubular geometries while our training and test datasets both use more complex geometries witnessed in the EM data. 

In case 2, PhySeg performs very well (dice score of 0.939) but our performance falls marginally (dice score 0.860 in case 1 versus 0.835 in case 2). PhySeg in this case benefits from the similarity of geometries in the training and test dataset. Our performance merits discussion here since our training dataset was far more complex and shape-wise the test dataset is a simpler subset. However, it is notable that our simulation dataset could not accommodate mitochondria of long lengths due to restricted computational resources while the test dataset did incorporate significantly lengthier mitochondrial geometries (as long as 5 $\upmu$m in comparison to our $\sim$1 $\upmu$m. 

In case 3 where manually annotated experimental test dataset is used, while our method suffers from small lengths of simulated mitochondria, PhySeg suffers from simulating simple geometry. This highlights the need of simulating realistic geometries as well as dimensions of mitochondria. However, in case 4 where we incorporate transfer learning on manually annotated experimental training dataset, our model quickly regains the performance and even exceeds its previous best performance (dice score 0.864 in case 4 versus 0.860 in case 1). On the other hand, the performance of PhySeg deteriorates from 0.889 in case 3 to 0.859 in case 4. This interesting shift in the performance trend is discussed here. Since our model already incorporates learning on complex shapes and is deficient primarily in the lengths of mitochondria, it quickly supplements in deficiency through transfer learning. On the other hand, PhySeg needs to augment its learning on simple geometries to the complex geometries present in the experimental dataset, which is a significantly more challenging endeavour than learning only the length aspect. 

We also include a brief comment on the performances of the two models across cases 1,2 and 4. Here, we exclude case 3 since case 4 directly builds upon case 3. PhySeg's best performance is in case 2 (0.939 dice score) and the worst performance is in case 1 (0.810 dice score). This indicates a large range in which the performance of PhySeg varies. Our best performance is in case 4 (0.864 dice score) and the worst performance is in case 2 (0.835 dice score). The range of performance of our model is relatively small, indicating a more robustness across various scenarios. 






\end{document}